\documentstyle[12pt]{article}
\newcommand{\lab}[1]{\label{#1}}

\newcommand{\rf}[1]{(\ref{#1})}

\newcommand\beq{\begin{equation}}
\newcommand\eeq{\end{equation}}
\newcommand\beqy{\begin{eqnarray}}
\newcommand\eeqy{\end{eqnarray}}
\newcommand\beqys{\begin{eqnarray*}}
\newcommand\eeqys{\end{eqnarray*}}

\newtheorem{Rem}{Remark}
\newtheorem{Con}{Convention}

\newtheorem{Thm}{Theorem}
\newtheorem{Cor}{Corollary}

\newcommand{\QED}{\begin{flushright}
Q.E.D.\end{flushright} }

\newcommand{\reali}{{\hbox{{\rm I}\kern-.2em\hbox{\rm R}}}}
\newcommand{\complessi}{{\ \hbox{{\rm I}\kern-.6em\hbox{\bf C}}}}

\newcommand{\ad}{{\,\hbox{\rm ad}}}
\newcommand{\Ad}{{\,\hbox{\rm Ad}}}

\newcommand{\Tr}{{{\rm Tr}\,}}

\newcommand{\lora}{{\longrightarrow}}

\newcommand{\Lra}{{\Leftrightarrow}}
\newcommand{\Ra}{{\Rightarrow}}
\newcommand{\rsa}{{\rightharpoonup}}

\newcommand{\dg}{{{\dag}}}
\newcommand\braket[2]{\left\langle{ {#1}\,,\,{#2} }\right\rangle}
\newcommand{\dphi}[1]{\frac{\overrightarrow\delta}{\delta\Phi^{#1}}}
\newcommand{\dphib}[1]{\frac{\overleftarrow\delta}{\delta\Phi^{#1}}}

\newcommand{\dphid}[1]{\frac{\overrightarrow\delta}{\delta
\Phi^\dg_{#1}}}
\newcommand{\dphidb}[1]{\frac{\overleftarrow\delta}{\delta
\Phi^\dg_{#1}}}

\newcommand{\antib}[2]{ \left({{#1}\,,\,{#2}}\right) }

\newcommand{\calA}{{\cal A}}
\newcommand{\calB}{{\cal B}}

\newcommand{\calG}{{\cal G}}

\newcommand{\calM}{{\cal M}}
\newcommand{\calN}{{\cal N}}
\newcommand{\calO}{{\cal O}}

\newcommand{\Gaff}{{\calG_{\rm aff}}}

\newcommand{\wcomm}[2]{{ \left[{ {#1}, {#2} }\right]}}

\newcommand{\adb}{{ {\rm Ad}_{b^{-1}} }}
\newcommand{\adg}{{ {\rm Ad}_{g^{-1}} }}

\newcommand{\SYM}{S_{\rm YM}}
\newcommand{\SYMfo}{S_{\rm YM'}}
\newcommand{\SBFYM}{S_{\rm BFYM}}
\newcommand{\PsiYM}{{\Psi_{\rm YM}}}
\newcommand{\ZYM}{{Z_{\rm YM}}}

\newcommand{\ZBFYM}{{Z_{\rm BFYM}}}
\newcommand{\sqtg}{{\frac1{\sqrt2\,g_{\rm YM}}\,}}
\newcommand{\gym}{g_{\rm YM} }
\newcommand{\Omplus}{{\Omega^{(2,+)}(M,\ad P)}}
\newcommand{\Omminus}{{\Omega^{(2,-)}(M,\ad P)}}
\newcommand{\tDelta}{{\widetilde\Delta}}
\newcommand{\tHarm}{{\widetilde{\rm Harm}}}
\newcommand{\tG}{{\widetilde G}}

\newcommand{\tpsi}{{\widetilde{\psi}}}
\newcommand{\tphi}{{\widetilde{\phi}}}
\newcommand{\bc}{{\overline c}}
\newcommand{\bxi}{{\overline\xi}}
\newcommand{\btpsi}{{\overline\tpsi}}
\newcommand{\btphione}{{\overline{\tphi_1}}}
\newcommand{\btphitwo}{{\overline{\tphi_2}}}
\newcommand{\htpsi}{{h_\tpsi}}
\newcommand{\hone}{{h_{\tphi_1}}}
\newcommand{\htwo}{{h_{\tphi_2}}}
\newcommand{\bk}{{\overline k}}
\newcommand{\bro}[1]{{\overline{r_1}^{#1}}}
\newcommand{\brt}[1]{{\overline{r_2}^{#1}}}
\newcommand{\hro}[1]{{h_{r_1}^{#1}}}
\newcommand{\hrt}[1]{{h_{r_2}^{#1}}}
\newcommand{\bchi}{{\overline\chi}}

\newcommand{\tb}[1]{{b^{#1}}}

\newcommand{\SBV}{S^{\rm BV}}
\newcommand{\bomega}{{\mbox{\boldmath $\omega$}}}

\newcommand{\tX}{\widetilde{X}}

\newcommand{\tZ}{\widetilde{Z}}

\newcommand{\lieg}{Lie(G)}
\newcommand\qq{\rm}
\newcommand\cmp[1]{{\qq Commun.\ Math.\ Phys.\ \bf #1}}
\newcommand\jmp[1]{{\qq J.\ Math.\ Phys.\ \bf #1}}
\newcommand\pl[1]{{\qq Phys.\ Lett.\ \bf #1}}
\newcommand\np[1]{{\qq Nucl.\ Phys.\ \bf #1}}
\newcommand\mpl[1]{{\qq Mod.\ Phys.\ Lett.\ \bf #1}}
\newcommand\pr[1]{{\qq Phys.\ Rev.\ \bf #1}}

\newcommand\lmp[1]{{\qq Lett.\ Math.\ Phys.\ \bf #1}}

\newcommand\cqg[1]{{\qq Class.\ Quant.\ Grav.\ \bf #1}}
\newcommand\prept[1]{{\qq Phys.\ Rept.\ \bf #1}}

\newcommand\anp[1]{{\qq Ann.\ Phys.\ \bf #1}}

\begin{document}

\begin{titlepage} 
\begin{center} 
{\large\bf Four-Dimensional Yang--Mills Theory \\
as a Deformation of Topological $BF$ Theory}
\\  
   
\vspace{2.0cm} 
{\bf A. S. Cattaneo},\footnote{
Lyman Laboratory of Physics,
Harvard University,
Cambridge, MA 02138, USA.
Supported by I.N.F.N.\  Grant No.\ 5565/95 and
DOE Grant No.\ DE-FG02-94ER25228, Amendment A003
}
{\bf P. Cotta-Ramusino},\footnote{
Dipartimento di Matematica, Universit\`a di Milano,
Via Saldini 50, 20133 Milano, ITALY}
\footnote{I.N.F.N. Sezione di Milano}
{\bf F. Fucito},\footnote{I.N.F.N. Sezione di Roma II,  
Via Della Ricerca Scientifica, 00133 Roma, ITALY}\\
{\bf M. Martellini},\hspace{1mm}$^{3}$
\footnote{Dipartimento di Fisica, Universit\`a  di Milano  
Via Celoria 16, 20133 Milano, ITALY}
\footnote{Landau Network  at ``Centro Volta,'' Como, ITALY}
{\bf M. Rinaldi},\footnote{
Dipartimento di Matematica, Universit\`a di Trieste, Piazzale Europa 1,
34100 Trieste, ITALY}
{\bf A. Tanzini},\hspace{1mm}$^{4}$
\footnote{Dipartimento di Fisica,  
Universit\`a di Roma II ``Tor Vergata," Italy}
{\bf M. Zeni}\hspace{1mm}$^3$
$^{5}$ 
\vskip 3.0cm 
{\large \bf Abstract}\\ 
\end{center} 
{The classical action for  
pure Yang--Mills gauge theory can be formulated as a deformation of  
the topological $BF$ theory where, beside the two-form field $B$, 
one has to add 
one extra-field $\eta$ given by a one-form
which transforms as the difference of two connections.
The ensuing action functional gives a theory that is both classically
and quantistically equivalent to the original Yang--Mills theory.
In order to prove such an equivalence,
it is shown that the dependency on the field $\eta$ can be 
gauged away completely. This
gives rise to a field theory that, for this reason, can be considered
as semi-topological or topological in some but not all the fields
of the theory. The symmetry group involved in this theory is an 
affine extension
of the tangent gauge group acting on the tangent bundle of the space of 
connections. A mathematical analysis of this group action and of the
relevant BRST complex is discussed in details.
}\vfill 
\end{titlepage} 

\section{Introduction}
Among the open problems of quantum Yang--Mills (YM) theory, there is 
certainly the absence of any proof of the property of confinement, 
which is observed in nature for systems supposedly described by a YM 
Lagrangian, and which is proved true only in lattice formulations of
the theory.

The non perturbative dynamics of gauge theories 
has been  discussed at length in the literature 
from different point of views. 
More recently, some of the authors of this paper observed \cite{CCGM,FMZ} 
that the presence of a two-form field in the first-order formulation of
YM theory might allow the construction of a surface observable 
that is related to `t~Hooft's magnetic order operator \cite{tH}. 
A preliminary description of such surface observables can be found in 
\cite{CM,Cth}; for a rigorous mathematical definition
of these observables (in the case of paths of paths), 
s.\ \cite{obs}.

The study of first-order YM theory was originally proposed in 
\cite{Halp}
as a way of taking into account strong-coupling effects (after some 
manipulations). For a discussion of this topic and its development, 
s.\ \cite{Rein} and references therein.

Another aspect of first-order YM theory pointed out in \cite{CCGM} is
its formal relationship with the topological $BF$ 
theory \cite{Schw,BlT}.\footnote{
For the study of $BF$ theory with observables, 
s.\ also \cite{BFobs}.
}\ \ This suggested
the possibility of finding Donaldson-like invariants \cite{Donaldson}
inside ordinary (not supersymmetric) YM theory with a mechanism
similar to that described in \cite{Witt}.

The relation between YM and $BF$ theory is actually more involved
because the latter has less physical degrees of freedom
than the former.

In this paper we establish the correct relation by using, as an 
intermediate step, a new theory---called $BF$-Yang--Mills (BFYM)
theory---which contains a new one-form field $\eta$ whose role is to provide
the missing degrees of freedom.\footnote{For an anticipation of
some results of this paper by some of the authors,
s.\ \cite{FMSTVZ}.}

The first aim of this paper is to prove the classical and quantum 
equivalence between YM and BFYM theory.  Cohomological
proofs of this were considered in \cite{Hen} and in
\cite{FMSTVZ}. In the present paper we give a different and
more explicit proof  
by fixing the gauge of the theory in
three different but equivalent ways, to which
we refer as to
the trivial, the covariant and the self-dual gauge fixing.

The reason for considering three different gauge fixings is that
each of them provides a different setting for perturbation theory:
In the trivial gauge, we have an expansion identical to that
found in first-order YM theory (s.\ \cite{beta}). 
In the covariant gauge, the perturbative
expansion around a flat connection can be organized by using the same
propagators as in the topological pure $BF$ theory (in the same gauge).
Finally, 
the perturbative expansion in the self-dual
gauge in a neighborhood of anti-self-dual
connections makes use of the propagators of the topological $BF$
theory with a cosmological term (in the same gauge).

The relation between BFYM and the $BF$ theories is then discussed in
a more formal way in the subsequent section by using the 
Batalin--Vilkovisky (BV) formalism \cite{BV} (which is a generalization
of the more familiar BRST formalism \cite{BRST}). 
In particular, we show that, after a {\em canonical transformation}, 
it is possible to perform
safely the limit in which the YM coupling constant vanishes and
obtain the pure $BF$ theory plus the (covariant) kinetic term for
the extra field $\eta$.

In the last part of the paper the geometrical aspects 
of BFYM are discussed.
The group of symmetries of the BFYM 
theory (for an extended discussion, s.\ \cite{CCR-tg})
turns out to be
an affine extension of the tangent gauge group.
The action of this group on the
space of fields is not free, but a BRST complex is obtainable 
directly from the action of the tangent gauge group on the space
of fields of the theory. 

The situation
has both similarities  and differences  with respect to the case
of topological gauge theories \cite{BS}. As in \cite{BS},
the BRST equations are obtained
as structure equations and Bianchi identities for the curvature of a suitable
connection on the space of fields; 
but, differently from \cite{BS}, the only symmetry for
the connection $A$ is the gauge invariance, as in the YM theory.
Hence the BFYM theory can be seen as ``semi-topological"
(or topological in the field $\eta$ and non-topological in the field $A$).

\section{Preliminaries}
\lab{sec-prem}
In this section we introduce YM theory both in its usual
second-order and in its first-order formulation.
We prove the equivalence of the two formulations
and discuss the problems related
to the weak-coupling perturbative expansion of the latter. 

Then we will raise the issue of the topological embedding for a gauge field 
theory and discuss some of its general properties.
It will be the aim of the following sections to prove that, through
the topological embedding, it is possible to define a
weak-coupling perturbative expansion of first-order YM theory
around the topological $BF$ theory.
\subsection{Second-order YM theory}
Let $P\to M$ be a principal $G$-bundle. The manifold $M$ is a
closed, simply connected, oriented four-manifold and $G$ is $SU(N)$ or,
more generally, a
simple compact Lie group. The standard  (second-order) YM action
is the following local functional 
\beq
\SYM[A] = \frac1{4\gym^2} \braket {F_A}{F_A},
\lab{YM}
\eeq
where $\gym$ is a real parameter (known as the YM coupling constant), and
$F_A$ is the curvature of the connection $A$. 
Here we consider the 
inner product $\braket\cdot\cdot$ defined on the space
$\Omega^*(M, \ad P)$ of forms on $M$ with values in the adjoint bundle
$\ad P$ and given by 
\beq
\braket\alpha\beta = \int_M \Tr (\alpha\wedge *\beta),
\lab{scalar}
\eeq
where $*$ is the Hodge operator on $M$. Even though the physical space--time 
is
Minkowskian, we assume we have performed 
a Wick rotation so that $M$ is a Riemannian manifold; viz., it
has a  {\em Euclidean} structure. 

The gauge group (or group of gauge 
transformations) is, by definition, the group $\calG$ of maps
$g:P\to G$ which are equivariant, i.e., which satisfy the 
equation $g(ph)={\rm Ad}_{h^{-1}} g(p)$ for any $p\in P$ and $h\in G$.
Locally elements of $\calG$ are represented by maps $M\to G$.
Under a gauge trasformation $g\in \calG$, the connection $A$ transforms as
\beq
A\to A^g =\adg A + g^{-1}dg,
\lab{Agauge}
\eeq
while any form $\psi\in \Omega^*(M, \ad P)$ transforms as
\[
\psi \to \adg \psi.\]
The Yang--Mills action is gauge-invariant, i.e., is invariant under the
action of $\calG$.
We denote the space of all connections by the symbol $\calA$. With some
restrictions, the group $\calG$ acts freely on $\calA$ and 
$\calA \to (\calA/\calG)$ is a principal 
$\calG$-bundle.\footnote{For instance we can consider only 
the space of irreducible connections 
and the action of the group obtained
by dividing  $\calG$ by its center. But in order to avoid cumbersome
notations we will keep on writing the quotient as 
$\calA/\calG$.
}

\subsection{Classical analysis}
A classical solution is a minimum of the action (modulo gauge
transformations), i.e., a solution of the equation
\beq
d_A^* F_A = 0.
\lab{YMeq}
\eeq
If we add a source $J(A)$ to the action, then the equations
of motion become $d_A^*F_A + 2\gym^2J' = 0$.

A particular class of solutions, 
is given by the (anti)self-dual connections,
i.e., connections whose curvature is (anti)self-dual (i.e., satisfies
the equation $F_A=\pm *F_A$).

We will denote by $\calM_{\rm YM}$ the moduli space of solutions
of the YM equations of motion modulo gauge transformations.

\subsection{Quantum analysis}
If we denote by $\calA$ the space of connections, then the partition
function is defined as
\beq
\ZYM = \int_{\calA/\calG} \exp(-\SYM).
\eeq

The way physicists deal with this quotient in the quantum
analysis is by introducing the BRST 
complex
\beq
\begin{array}{c | ccc}
 & -1 & 0 & 1\\ \hline
1 & & A & \\
0 & \bc & h_c & c
\end{array}
\lab{BRSTc-YM}
\eeq
(where each row has the same form-degree and each column has 
the same  ghost-number), and the BRST transformations:
\beq
sA = d_Ac,\quad sc = -\frac12[c,c],
\lab{sAc}
\eeq
\beq
s\bc = h_c,\quad s h_c = 0.
\lab{sch}
\eeq
Notice that the first equation is just the infinitesimal
version of the action \rf{Agauge} of $\calG$
on $\calA$, while the
second is the Maurer--Cartan equation for the group $\calG$.

A (local) section $\calA/\calG\to \calA$  
is  chosen by introducing a gauge fixing, i.e., by imposing a condition like, 
e.g.,
$d_{A_0}^*(A-A_0)=0$ for $A$ belonging to a suitable neighborhood of
$A_0$. Correspondingly, one introduces a gauge-fixing
fermion $\PsiYM$, i.e., a local functional of ghost number $-1$ given 
by\footnote{
This way of implementing a gauge fixing is known as the {\em Landau
gauge}. A more general gauge-fixing fermion implementing the same
conditions is obtained by the replacement $\PsiYM\to
\PsiYM + \lambda\braket\bc{h_c}$, where $\lambda$ is a free parameter.
By integrating out $h_c$ and setting $\lambda=1$, one recovers the 
{\em Feynman gauge}.

In this paper we will consider only Landau gauges.
}
\beq
\PsiYM = \braket{\bc}{d_{A_0}^*(A-A_0)},
\lab{PsiYM}
\eeq
where $A_0$ is a background connection. In perturbative calculations,
we work in a neighborhood of a critical solution; 
i.e., we assume that $A_0$  is a solution of
the classical equations of motion.

The original action is then replaced by
\beq
\SYM^{\rm g.f.} = \SYM + is\PsiYM,
\eeq
and the functional 
integration is performed over the vector spaces to which the fields
of the BRST complex belong (notice that the integration over the affine
space of connections is replaced by an integration over the vector 
space $\Omega^1(M, \ad P)$ to which $A-A_0$ belongs).

To perform computations, it is also useful to assign each field
a canonical (scaling) dimension so that the gauge-fixed action has 
dimension zero. Since the derivative, the volume integration 
and the BRST operator
have respectively dimension $1$, $-4$ and $0$, we get the following
table:
\begin{description}
\item{\sf Dimension 0:} $c$.
\item{\sf Dimension 1:} $A$.
\item{\sf Dimension 2:} $\bc$, $h_c$.
\end{description}
Notice that the coupling constant $\gym$ has dimension 0.

\paragraph{Perturbative expansion}
The perturbative expansion of YM theory around a critical connection
$A_0$ is performed by setting
\beq
A = A_0 + \sqrt2\,\gym\,\alpha.
\eeq
By also using the gauge fixing condition, the quadratic part
of the action then reads
\beq
\frac12\braket\alpha{\check\Delta_{A_0}\alpha} +
i\braket{h_c}{d_{A_0}^*\alpha} -i\braket\bc{\Delta_{A_0}c},
\lab{YMq}
\eeq
where
\beq
\check\Delta_{A_0} = \Delta_{A_0} + *[*F_{A_0},\ ],
\lab{cDelta}
\eeq
and $\Delta_{A_0}$ is the covariant Laplace operator.
The $\alpha\alpha$-propagator is given by the inverse of
$\check\Delta_{A_0}$.

\subsection{The first-order formulation of YM theory}
\lab{sec-foYM}
Now we consider  the local functional
\beq
\SYMfo[A,E] = i\braket E{*F_A} + \gym^2\braket EE,
\lab{YM'}
\eeq
where $F_A$ is the curvature of the connection $A$  and
$E\in\calB\equiv\Omega^2(M, \ad P)$. As for the canonical dimension,
$E$ is assigned dimension $2$.

The first-order YM theory---which we will prove in a moment to be
equivalent to YM theory
both at the classical and at the quantum level---is
particularly interesting because of the new independent field $E$ which
allows the introduction of new observables which depend on loops of paths
on $M$ (or on the spanned surfaces) 
and could not be defined in ordinary YM theory \cite{obs}.

The only symmetry of the theory corresponding to 
\rf{YM'} is the gauge symmetry.
It acts on the space of fields
$\calA\times\calB$ as in \rf{Agauge} plus
\beq
E\to\adg E.
\lab{Egauge}
\eeq
The group $\calG$ acts freely on the manifold $\calA\times\calB$, 
which becomes a principal $\calG$-bundle with the projection 
$(\calA\times\calB)\to\calA$ being a bundle-morphism.
As a consequence, the gauge fixing on $A$ is enough to completely fix
the first-order formulation of YM theory.

\paragraph{Remark} The presence of an $i$ in the action \rf{YM'} may
look odd but is necessary since the $EF$ term is not positive 
definite. Notice that without the Wick rotation (thus on a Minkowskian 
manifold), the factor $i$ would disappear from the action.

\subsubsection{Classical equivalence}
The critical points (which are not minima) of the action \rf{YM'}
correspond to the solutions of the following equations of motion:
\beq
\begin{array}{lcl}
i*F_A + 2\gym^2 E &=& 0,\\
i*d_AE &=& 0.
\end{array}
\lab{YMfoeom}
\eeq
By applying the operator $*d_A$ to the first equation,
we see that, because of the second equation, 
$A$ must solve the YM equation \rf{YMeq}. By the first equation
we then see that $E$ must
be equal to $-(i/2\gym^2)*F_A$. 
Thus, the space of solutions
of first-order YM theory is in a one-to-one correspondence with
the space of solutions of second-order YM theory. Moreover, this
correspondence is preserved by gauge equivalence. Therefore, the
moduli spaces of the two theories are the same.

The presence of an $i$ is a bit disturbing since it requires
an imaginary solution for $E$. Moreover, it could seem
that the factor $i$ does not play any role in the classical equations. 
However, if we
add a source $J(A)$ to the action \rf{YM'}, the second equation
is replaced by $i*d_AE+J'=0$. The application of $*d_A$ to the
first equation gives the correct answer $d_A^*F_A+2\gym^2J'=0$ just
because of the $i$ factor. 
Observe that if we were working on a Minkowski space,
then we would get the correct answer by removing the factor $i$ (this
is because $*^2$ keeps track of the signature of the metric).

\subsubsection{Quantum equivalence}
It is not difficult to see that a Gaussian integration over
$E$ yields
\beq
\int_{(\calA\times\calB)/\calG} \exp(-\SYMfo[A,E])
\ \calO[A]
\propto \int_{\calA/\calG} \exp(-\SYM[A])\ \calO[A],
\lab{YM'YM}
\eeq
where $\calO$ is any gauge-invariant observable for YM theory.

Notice that the proportionality constant depends on $\gym$ because
of the determinant coming from the $E$-integration. This dependence
can be removed if one defines the functional measure for $E$ as already
containing this factor. Anyhow, this constant factor is irrelevant since
computing vacuum expectation values involves a ratio,
and we will not take care of it.


More explicitly, the integration can be performed by introducing
the BRST complex and the BRST transformations \rf{sAc} and \rf{sch}
plus
\beq
sE = [E,c],
\lab{sE}
\eeq
which corresponds to the infinitesimal version of the gauge
transformation \rf{Egauge}. 

Notice again that the presence of the $i$ factor in the action is essential
to make the Gaussian integration meaningful and to get the correct
answer [instead of $\exp(+\SYM)$].

Finally, it is important to notice that $\calB$ is a vector space, so
{\em it is not necessary to fix a background solution $E_0$ to perform
the integration; therefore, the equivalence between first- and 
second-order YM theories is non-perturbative}. However, one might also
decide to fix a background solutions $A_0$ of YM equations and a
background field $E_0=-(i/2\gym^2)*F_{A_0}$ and integrate in the variable
$E-E_0$; the result will be an equivalence with YM theory expanded
around the same $A_0$.

\subsubsection{The perturbative expansion}
The exact computation of a functional integral is a formidable task.
Usually one considers a perturbative expansion around a classical
solution. YM theory can be computed as an asymptotic series in
$\gym$ (i.e., in weak coupling)
after defining the integration variable as 
$\alpha=(A-A_0)/(\sqrt2\,\gym)$.
In the first-order formulation, perturbation theory requires 
choosing a background for $E$ as well and introducing the integration
variable $\beta=\sqrt2\,\gym\,(E-E_0)$. The action will then contain
the quadratic part
\[
\braket\beta{*d_{A_0}\alpha}+
\braket{2\gym^2 E_0}{*(\alpha\wedge\alpha)}
+\frac12\braket\beta\beta + \mbox{gauge fixing}
\]
plus terms of order $\gym$ [notice that $\gym^2E_0=O(1)$]. 
{}From the quadratic part one reads the propagators and the Feynman
rules leading to the usual ultraviolet behaviour \cite{beta}.

Another possibility, which is interesting exploring,
is to consider the term $\braket\beta\beta$ as a perturbation.
In this way the propagators will resemble those of the topological
pure $BF$ theory defined by the action
\beq
S_{BF} = i \braket B{*F_A},
\eeq
where the field $B$ is just a new name for our $E$.
Unfortunately, however, the operator
acting on $(\alpha,\beta)$ in this scheme is not invertible
since its kernel includes
any pair $(0,\beta)$ with $\beta\in \ker(d_{A_0})$. 
In the pure $BF$ theory there is no problem since
the theory itself has a larger set of symmetries (s.\ Sec.\ 
\ref{sec-BV}) and, therefore, an additional gauge fixing
is required.

Another way of seeing our problem is the following: The pure $BF$
theory appears formally as the $\gym^2\to0$ limit of the first-order YM
theory (after renaming $E$ as $B$). However, the number of
degrees of freedom are reduced in this limit and this shows that
the limit is ill-defined.

The purpose of the next sections is to show that is
possible to restore the missing degrees of freedom by introducing an extra
field, and that this makes the above limit meaningful.
The mechanism that will allow us to do so is the so-called
``topological embedding."

\paragraph{Remark} 
Notice that in the first-order YM theory
one can define also a strong-coupling expansion (i.e.,
an asymptotic expansion in $1/\gym$) after integrating out the 
connection
in \rf{YM'} (notice that this integration is Gaussian). For details,
s.\ \cite{Halp,Rein}.

\subsection{The ``topological embedding"}

The so-called topological embedding refers to 
the idea of  ``embedding" a topological into a 
physical theory.

The way we discuss such a scheme is partly related to the
arguments presented in \cite{Ans}.
The basic idea is to consider an action $S[A]$ (or, more generally, an
action $S[A,E]$) as a functional of an auxiliary field $\eta$ as well.
One then writes $S[A,\eta]$, but it is understood that $\delta 
S/\delta\eta=0$.

Of course, this gives the theory a huge set of symmetries, viz., all
possible shifts of $\eta$, so that $\eta$ has no physical degrees
of freedom. This is similar to what happens in the 
topological field theories of the so-called cohomological (or Witten) type
where all fields are subject to such a symmetry. One might also
speak of semi-topological theory since it is topological, in the
previous sense, only in some field directions.

In our case the new field $\eta$
belongs to $\Omega^1(M, \ad P)$, so that the pair $(A,\eta)$ is an
element of the tangent bundle $T\calA$. The Lie group $T\calG$, which
can be represented as the semi-direct product of $\calG$ and
the abelian group $\Omega^0(M,\ad P)$, has a natural
action on $T\calA$ given by
\beq
(A,\eta) \rightarrow (A^g, \adg \eta + d_{A^g}\zeta),
\eeq
where $A^g$ is defined in \rf{Agauge} and $(g,\zeta)\in T\calG$ [here
$\zeta\in \Omega^0(M, \ad P)$].

It is convenient to combine the $T\calG$-transformation with the
translations acting on the field 
$\eta$ under which the theory is invariant; viz., we write
the transformation of $(A,\eta)$ as
\beq
(A,\eta)\to \left(A^g,\adg \eta + d_{A^g}\zeta -\tau\right),
\lab{etagauge}
\eeq
where $\tau\in\Omega^1(M, \ad P)$ represents the translation.
In this way we can write the group of the symmetries  of the theory as the
semidirect product of $T\calG$ with $\Omega^1(M, \ad P)$.
In the following we will denote this group by $\Gaff$.

Unfortunately, the action of $\Gaff$ given by \rf{etagauge} is not free.
However, as will be shown in detail in subsec.\ \ref{subsec-restr}, 
this problem 
can be successfully dealt with by considering 
the BRST complex defined by
\beq
s\eta = [\eta,c] + d_A\xi - \tpsi,
\lab{seta}
\eeq
where $\xi$ and $\tilde\psi$ are new ghosts. For the BRST operator
to be nilpotent they must obey the 
transformation rules
\beq
s\xi=-[\xi,c]+\tphi,\quad s\tpsi=-[\tpsi,c]+d_A\tphi,
\lab{sxitpsi}
\eeq
where $\tphi$ is a ghost-for-ghost (i.e., has ghost number 2)
which transforms as
\beq
s\tphi=[\tphi,c].
\lab{stphi}
\eeq

For further details on the space $T\calA$, on its symmetries and
on the implementation of the BRST procedure, s.\ Sec.\ \ref{sec-geo}
and Ref.\ \cite{CCR-tg}.

The quantization of such a theory requires a gauge fixing for this
topological symmetry as well. The apparently trivial operation of
adding a new field on which the theory does not depend and then gauging
it away can have some interesting consequences:
\begin{enumerate}
\item A trivial gauge fixing
for $\eta$ (i.e., setting $\eta=0$) is always available
but if a non-trivial  gauge fixing for $\eta$ is chosen, this 
may introduce a non-trivial measure on the moduli space. Heuristically
we have
\beq
\int_{T\calA/\Gaff} \exp(-S[A])
= \int_{\calA/\calG} \exp(-S[A])\ \mu[A,M],
\lab{functeqn}
\eeq
where\footnote{\lab{foot-mani} The quotient space $T\calA/\Gaff$ is 
not a manifold
since, as we discussed before, the action of $\Gaff$ is not free.
However, a 
BRST structure and the relevant quantization
for the $\Gaff$ symmetry is available.
This is the reason why we can still write, at the heuristic level,
the identity \rf{functeqn} between functional integrals.}
$\mu$ is the outcome of the $\eta$-integration and depends on
$A$ and on $M$ and its metric structure through
the chosen gauge fixing. Since one expects quantization not to depend
on small deformations of the gauge fixing (in particular on small
deformations of the
metric of $M$),
one can argue that $\mu[A,M]$
is a (possibly trivial) measure which, once integrated over the space
of critical solutions modulo gauge-transformations (moduli space),
gives a smooth invariant of $M$.

In particular
we may choose as non-trivial gauge fixing an incomplete one
which leaves only a finite number of symmetries, and then
introduce a
``topological observable" 
as a volume form.\footnote{
These are essentially the motivations of the approach of
\cite{Ans} where
the following version of the YM action
\[
S[A_0,A_q] = \frac1{4\gym^2} \braket{F_{A_0+\gym A_q}}{F_{A_0+\gym A_q}}
\]
is considered.
This is equivalent to choosing $S[A,\eta]=(1/4\gym^2)
\braket{F_A}{F_A}$ by the change of variable (with constant Jacobian)
\[
A_0 = A-\eta,\quad A_q = \frac1{\gym}\,\eta.
\]
Moreover, if one also considers the following change of ghost variables
(with constant Jacobian)
\[
\begin{array}{lclclcl}
C_0 &=& c-\xi, &\quad&C_q&=&\frac1{\gym}\,\xi,\\
\psi_0&=&\tpsi-[\eta,\xi],&\quad&\phi_0&=&\frac12[\xi,\xi]-\tphi,
\end{array}
\]
then the BRST transformations \rf{sAc},
\rf{seta}, \rf{sxitpsi} and \rf{stphi} become
\[
\begin{array}{lclclcl}
sA_0&=&\psi_0+d_{A_0}C_0, &\quad& sC_0&=&\phi_0-\frac12[C_0,C_0],\\
s\psi_0&=&-d_{A_0}\phi_0-[\psi_0,C_0],&\quad&s\phi_0&=&[\phi_0,C_0],\\
sA_q&=&-\frac1{\gym}\psi_0+d_{A_0+\gym A_q}C_q+[A_q,C_0],&\quad&
sC_q&=&-\frac1{\gym}\phi_0-[C_0,C_q]-\frac \gym2[C_q,C_q]
\end{array}
\]
(which correspond to the BRST trasformation listed in \cite{Ans}). 
In this way one recognizes
the topological set of transformations for $A_0$ which is 
later reinterpreted as the background connection.
}
\item When this procedure is applied the first-order formulation
of YM theory, the limit $\gym^2\to0$ becomes meaningful, as we will see
in the next sections. Moreover, a weak-coupling perturbation theory
with the propagators of one of the topological $BF$ theories becomes available.
\end{enumerate}

Before discussing the last point, it is better to rewrite the action
$\SYMfo[A,E,\eta]$ of 
(extended) first-order YM theory by making the change
of variables
\beq
B = E + \sqtg d_A\eta.
\lab{BE}
\eeq
This yields the action
\beq
\SBFYM[A,B,\eta] = i \braket B{*F_A} + 
\gym^2\braket{B-\sqtg d_A\eta}{B-\sqtg d_A\eta},
\lab{SBFYM}
\eeq
which we will call BFYM theory since it is related to both the YM theory,
after integrating out $B$ and $\eta$, and to the
$BF$ theory, in the limit $\gym^2\to0$;
in particular, this limit yields
\beq
\SBFYM[A,B,\eta]\stackrel{\gym^2\to0}\sim i \braket B{*F_A} +
\frac12\braket{d_A\eta}{d_A\eta},
\lab{SBFYM0}
\eeq
where we recognize the $BF$ theory plus a non-topological term that
restores the degrees of freedom of YM theory. Notice that
the presence of $1/(\sqrt2\, \gym)$ in \rf{BE} is designed so as to
give the kinetic term for $\eta$ the correct normalization.

In the next section we will reconsider the equivalence of the BFYM and
YM theories and prove it explicitly by using three different
gauge fixings.

In section \ref{sec-BV} we will discuss the limit $\gym^2\to0$
and show that it is well-defined in the present context.

\section{The BFYM theory}
\lab{sec-BFYM}
In this section we discuss the theory described by the action
\rf{SBFYM}, i.e.,
\[
\SBFYM[A,B,\eta] = i \braket B{*F_A} + 
\gym^2\braket{B-\sqtg d_A\eta}{B-\sqtg d_A\eta}.
\]

First we consider the equations of motion. They can be obtained directly
by looking at the stationary points of \rf{SBFYM} or from the 
equations of motion \rf{YMfoeom} of the first-order YM theory 
together with the change of variables \rf{BE}. In any case, the 
equations of motion can be written, after a little algebra,  as
\beq
\begin{array}{lcl}
i*F_A + 2\gym^2 B -\sqrt2\,\gym\,d_A\eta &=&0,\\
d_A^*F_A &=& 0,\\
d_A F_A &=& 0.
\end{array}
\lab{eom}
\eeq
Notice
that the third equation is just the Bianchi identity.

Then we come to the symmetries. They can be obtained starting
from the symmetries \rf{Agauge} and \rf{Egauge} of the first-order
YM theory and from the topological symmetry \rf{etagauge} for $\eta$
together with the change of variables \rf{BE}. Explicitly, we have
\beq
\begin{array}{lcl}
A &\to& A^g =\adg A + g^{-1}dg,\\
\eta&\to& \adg \eta + d_{A^g}\zeta -\sqrt2\,\gym\,\tau,\\
B &\to& \adg B + \sqtg [F_{A^g},\zeta] - d_{A^g}\tau,
\end{array}
\lab{ABetagauge}
\eeq
with $(g,\zeta,\tau)\in \Gaff$. The action of $\Gaff$ on the space
of triples $(A,\eta,B)$ is again not a free action, but again, as it 
will be shown in a moment,
a BRST 
complex and the relevant quantization are available.

Notice that we have rescaled 
$\tau\to\sqrt2\,\gym\,\tau$ so as to see the 
shift on $\eta$ as
a perturbation of the  tangent group action; this is consistent with
the limit $\gym^2\to0$ in \rf{SBFYM0}. 
We will discuss this issue better in the next section, where we
also explain why the $B$-transformation becomes singular in this limit.
Notice that for the computations considered in this section all these
rescalings are irrelevant.

A further remark concerns the geometric interpretation of the field $B$:
since it transforms as $d_A\eta$, it is natural to see it as an element
of a the tangent space $T_{F_A}\calB$ and not of $\calB$. 

To quantize the theory we have to describe
the BRST symmetry. Again the BRST transformations
for BFYM theory can be obtained from \rf{sAc}, \rf{sE}, \rf{seta},
\rf{sxitpsi}, \rf{stphi} and \rf{BE}. Explicitly they read
\beq
\begin{array}{lclclcl}
sA &=& d_Ac, &\quad& sc &=& -\frac12[c,c],\\
s\eta &=& [\eta,c] + d_A\xi - \sqrt2\, \gym\,\tpsi, &\quad&
s\xi &=&-[\xi,c]+\sqrt2\, \gym\,\tphi,\\
sB &=& [B,c] + \sqtg [F_A,\xi] - d_A\tpsi, & & & & \\
s\tpsi&=&-[\tpsi,c]+d_A\tphi, &\quad& 
s\tphi &=& [\tphi,c].
\end{array}
\lab{BRSTBFYM}
\eeq
where $c$ and $\xi$ have ghost number 1 and belong to $\Omega^0(M, \ad P)$,
$\tpsi\in\Omega^1(M, \ad P)$ and has ghost number 1, and $\tphi\in
\Omega^0(M, \ad P)$ and has ghost number 2. Notice that we have
rescaled $(\tpsi,\tphi)\to\sqrt2\, \gym(\tpsi,\tphi)$ so as to
see the shifts as perturbations of the $T\calG$ transformations
on $\eta$ and $\xi$.

To study the theory, both at the classical and at the quantum level, 
we have to fix the symmetries \rf{ABetagauge}. 
After having done this,
our first aim will be to prove that the gauge-fixed BFYM and YM theories
are classically equivalent, i.e., that their moduli spaces are
in one-to-one correspondence with each other.
Our second aim will be to prove the quantum equivalence, i.e.,
\beq
\int_{(T\calA\times T_{F_A}\calB)/\Gaff}
\exp(-\SBFYM[A,\eta,B])\ \calO[A]\propto
\int_{\calA/\calG} \exp(-\SYM[A])\ \calO[A].
\lab{BFYMYM}
\eeq
As in \rf{YM'YM}, the proportionality constant will depend on $\gym$ but
will not affect the vacuum expectation values, and we will not
take care of it.

Notice that, as it was for the case in \rf{functeqn},
the quotient $(T\calA\times T_{F_A}\calB)/\Gaff$ is not a manifold
since the action of $\Gaff$ is not free. The same argument of footnote
\ref{foot-mani} applies here and a detailed discussion 
on how to deal with the non-freedom
of these group action is considered in subsec.\ \ref{subsec-Bfield}.

The formal computation of the functional integral can be performed
after 
choosing a gauge fixing. In general, whenever we verify that
some conditions are a gauge fixing (at least in a neighborhood of
critical solutions), we expect the equivalence to be realized
(in that neighborhood); for we can always go back to the
variables $A,\eta,E$ by \rf{BE} and perform the Gaussian 
$E$-integration. The $\eta$-integration should give at most
some topological contributions since $\eta$ appears only
in $s$-exact terms now. However, the change of variables \rf{BE}
becomes singular as $\gym\to0$, so we prefer to work the equivalence out
by using the variables $A,\eta,B$.

We will consider three different gauge fixings which we call the 
{\em trivial}, the {\em covariant} and the {\em self-dual} gauge 
fixings.
The last two of them
will be dealt with in the next two subsections, and the
conditions under which classical and quantum equivalence are true
will be discussed. 
As for the trivial gauge fixing, characterized by the condition $\eta=0$
plus a gauge-fixing condition on $A$, we see immediately that BFYM 
theory turns out to be
equivalent to the first-order formulation of YM theory which,
as we proved in the previous section, is equivalent to the
second-order formulation. 

The other two gauge fixings are equivalent to the trivial one (when 
they are defined), so we can be sure of the 
equivalence between BFYM and YM theory in any of these gauges without 
any further computation.
However, we prefer to check the equivalence
explicitly, for this treatment also produces the correct framework to
consider perturbation theory around $BF$ theories. 

Obviously a weak-coupling expansion as in first-order YM theory is 
always possible, and this is the only possibility in the trivial gauge.
In the covariant gauge, however, we will show that
perturbation theory around a flat connection
can be organized in a different way so that the $AB$-sector and the 
$\eta$-sector of the theory decouple in the unperturbed action and the
$AB$-propagator turns out to be the propagator of the topological 
pure $BF$
theory (in the covariant gauge). Finally, in the self-dual gauge, 
perturbation theory around an anti-self-dual non-trivial connection (the 
only kind of connection around which this gauge is well defined) can
again be organized in such a way that the $AB$-sector and the 
$\eta$-sector decouple in the unperturbed action; moreover, the 
propagators in the $AB$-sector are recognized as those of the 
topological $BF$ theory with a cosmological term (in the self-dual 
gauge).

\subsection{The covariant gauge fixing}

The covariant gauge fixing, which will be discussed explicitly in
subsec.\ \ref{subsec-gf}, is characterized by a gauge-fixing condition
on $A$ together with
\beq
d_A^*\eta=0,\quad \eta\perp{\rm Harm}^1_A(M, \ad P),\quad
d_A^*B\in d_A\Omega^0(M, \ad P),
\lab{covetaB}
\eeq
where
\beq
{\rm Harm}^k_A(M, \ad P) \equiv \{\omega\in\Omega^k(M, \ad P)
\ |\ \Delta_A\omega=0\},
\eeq
and
\beq
\Delta_A \equiv d_A^*d_A+d_Ad_A^* :\Omega^*(M, \ad P)\to\Omega^*(M, \ad P).
\eeq
Notice that if $b^1[A]={\rm dim}\;{\rm Harm}^1_A(M,\ad P)$ is not constant on
the whole space $\calA$, the covariant gauge fixing is consistently 
defined only in those open regions where it is constant. In particular,
we will denote by $\calN$ the open neighborhood of the space of connections
where this is true (in particular cases, $\calN$ may be the whole
$\calA$).

By consistency, on the shift $\tau$
in \rf{ABetagauge} we must
impose the same conditions as those which fix the $T\calG$
symmetry on $\eta$, viz.,
\[
d_A^*\tau=0,\quad \tau\perp{\rm Harm}^1_A(M, \ad P).
\]
Similarly, in the context of BRST quantization, the ghost $\tpsi$ is
subject to the same conditions
\beq
d_A^*\tpsi=0,\quad \tpsi\perp{\rm Harm}^1_A(M, \ad P).
\lab{covtpsi}
\eeq

Since we have
${\rm Harm}_A^0(M, \ad P)=\{0\}$ (which is a consequence
of taking $A$ irreducible), this is actually a gauge 
fixing. 
Notice that there is a set of interpolating (complete) gauge fixings
between \rf{covetaB} and 
the trivial gauge fixing, $\eta=0$, which can also be written as
\[
d_A^*\eta=0,\quad
\eta\in d_A\Omega^0(M, \ad P).
\]
The interpolating gauge fixings are then given by
\[
\lambda d_A^* B + (1-\lambda)
\eta\in d_A\Omega^0(M, \ad P),
\]
with $\lambda\in[0,1]$. 

One might also choose $\lambda$ to be smooth but not constant on $\calA$. 
In particular, one could choose $\lambda$ to be constant and equal to 1
in an open neighborhood of the space of critical connections contained
in the neighborhood $\calN$, and constant and equal to 0 outside 
$\calN$. In this way one would obtain a gauge fixing that is defined on the 
whole space $\calA$ and restricts to the covariant gauge fixing close to the 
critical connections.

\subsubsection{Classical equivalence}
Consider the equations of motion \rf{eom}. The second and the third
tell us that $A$ is a solution of the YM equations. The first implies that
\[
d_A^*(2\gym^2 B -\sqrt2\,\gym\,d_A\eta) =0,
\]
so that
\[
\braket{d_A\eta}{2\gym^2 B -\sqrt2\,\gym\,d_A\eta}=0.
\]
On the other hand, the gauge-fixing conditions \rf{covetaB} 
imply that
\[
\braket{d_A\eta}B = 0.
\]
So we conclude that
\[
||d_A\eta||^2=0.
\]
By the positivity of the norm (remember that we are in a 
Riemannian manifold)
we get then $d_A\eta=0$.
Since the gauge fixing also imposes $d_A^*\eta=0$
and requires $\eta$ not to be harmonic, we conclude that
\beq
\eta=0.
\lab{covetasol}
\eeq
Finally, inserting this result in \rf{eom} yields
\beq
B = -\frac i{2\gym^2} *F_A.
\lab{covBsol}
\eeq

Therefore, we have shown that a solution $A$ of the YM equations completely
determines a solution of BFYM equations in the covariant gauge fixing.

Notice that this solution coincides with that obtained with
the trivial gauge fixing.

\subsubsection{Quantum equivalence}
To implement the covariant gauge fixing in the BRST formalism, we have
first to introduce the full BRST complex which generalizes 
\rf{BRSTc-YM}. It is useful to organize all the fields in the following
tables where each row has the same form-degree and each column has 
the same  ghost-number:
\beq
\begin{array}{c|ccc}
 & -1 & 0 & 1\\ \hline
1 & & (A,\eta) & \\
0 & (\bc,\bxi) & (h_c,h_\xi) & (c,\xi)
\end{array}
\lab{BRSTc-Aeta}
\eeq
\beq
\begin{array}{c|ccccc}
  & -2 & -1 & 0 & 1 & 2 \\ \hline
2 &    &    & B &   &   \\
1 &    & \btpsi & \htpsi & \tpsi & \\
0 & \btphione & \hone & \btphitwo & \htwo & \tphi
\end{array}
\lab{BRSTc-Btpsitphi}
\eeq
The BRST transformations are given by \rf{BRSTBFYM} together with
\beq
\begin{array}{lclclclclclclcl}
s\bc&=&h_c,&\quad&sh_c&=&0,&\quad&s\bxi&=&h_\xi,&\quad&sh_\xi&=&0,\\
s\btpsi&=&\htpsi,&\quad&s\htpsi&=&0, \\
s\btphione&=&\hone,&\quad&s\hone&=&0,&\quad&
s\btphitwo&=&\htwo,&\quad&s\htwo&=&0.
\end{array}
\eeq

If harmonic one-forms are present, in order to implement the covariant
gauge fixing, \rf{covetaB} and \rf{covtpsi},
it is better to rewrite the BRST
transformations for $\eta$ and $\tpsi$ displaying the harmonic
contribution. 

First we take an orthogonal basis $\omega_i[A]$ of
${\rm Harm}^1_A(M, \ad P)$, with $i=1,\ldots,\tb 1[A]=
\dim{\rm Harm}^1_A(M, \ad P)$. 
To be consistent with the scaling dimensions, we normalize
this basis as
\beq
\braket{\omega_i[A]}{\omega_j[A]} = \delta_{ij}\,\sqrt V,
\lab{omnorm}
\eeq
where $V$ is the volume of the manifold $M$.

As a consequence of the
fact that $\omega_i[A^g]=\adg\omega_i[A]$, we get
the BRST transformation rule
\beq
s\omega_i[A]=[\omega_i[A],c].
\eeq

Then we add a family of constant ghosts $k^i$ and
$r^i$ (respectively of ghost number 1 and 2) and BRST transformation
rules
\beq
sk^i=\sqrt2\,\gym\,r^i,\quad sr^i=0.
\lab{skr}
\eeq
Finally, we rewrite the BRST transformations for $\eta$ and $\tpsi$ as
\beq
\begin{array}{lcl}
s\eta &=& [\eta,c] + d_A\xi - \sqrt2\, \gym\,\tpsi + k^i\omega_i[A],\\ 
s\tpsi&=&-[\tpsi,c]+d_A\tphi +r^i\omega_i[A], 
\lab{etatpsih}
\end{array}
\eeq
where a sum over repeated indices is understood. It is easily verified
that the BRST operator is still nilpotent.

To implement the gauge fixing, we have then to build the BRST complex,
i.e., add to \rf{BRSTc-Aeta} and \rf{BRSTc-Btpsitphi} the 
following table:
\beq
\begin{array}{ccccc}
-2 & -1 & 0 & 1 & 2 \\ \hline
 & \bk^i & h_k^i & k^i \\
\bro i & \hro i & \brt i & \hrt i & r^i
\end{array}
\eeq
where each column has 
the displayed ghost-number.
We conclude
by giving the last BRST transformations, viz.,
\beq
\begin{array}{lclclclclclclcl}
s\bk^i &=& h_k^i, & \quad & sh_k^i &=& 0,\\
s\bro i &=& \hro i, &\quad & s\hro i &=& 0, &\quad&
s\brt i &=& \hrt i, &\quad & s\hrt i &=& 0.
\lab{skrb}
\end{array}
\eeq

Now we are in a position to write down the gauge-fixing fermion
that implements the conditions \rf{covetaB} and \rf{covtpsi}:
\beq
\Psi \begin{array}[t]{cl} 
= & \PsiYM + \\
+ & \braket\bxi{d_A^*\eta} + \bk^i\braket{\omega_i[A]}\eta +\\
+ & \braket\btphione{d_A^*\tpsi} 
+ \bro i\braket{\omega_i[A]}\tpsi +\\
+ & \braket\btpsi{d_A^*B+d_A\btphitwo+\brt i\,\omega_i[A]},
\end{array}
\lab{Psi}
\eeq
where $\PsiYM$ is a gauge-fixing fermion for YM theory like, e.g., 
in \rf{PsiYM}. Notice that both $d_A^*B$ and $d_A\btphitwo$
are in the orthogonal complement of ${\rm Harm}^1_A(M, \ad P)$;
thus, to implement the second gauge-fixing condition in
\rf{covetaB}, we must take $\btpsi$ in this orthogonal complement
as well. This is accomplished by the last term in \rf{Psi}.

The gauge-fixed action will then read
\beq
\SBFYM^{\rm g.f.} = \SBFYM + is\Psi.
\eeq

Notice the double role played here by $\btpsi$,
$\btphitwo$ and $\brt i$:
On the one hand, we can see $\btpsi$ as the antighost 
orthogonal to the harmonic forms
that allows
an explicit implementation of the gauge-fixing condition for $B$, viz.,
\beq
d_A^*B+d_A\btphitwo= 0;
\lab{covB}
\eeq 
on the other hand, we can see $\btphitwo$ 
and $\brt i$ as antighosts that implement on $\btpsi$ the same
conditions as those satisfied by $\tpsi$, viz., 
\beq
d_A^*\btpsi=0, \quad \btpsi\perp{\rm Harm}^1_A(M, \ad P).
\lab{covbtpsi}
\eeq

As in the case of YM theory, it is useful to assign a canonical
dimension to all the fields in such a way that the gauge-fixed
action, the derivative, the volume integration 
and the BRST operator have,
respectively, dimensions $0$, $1$, $-4$ and $0$. Therefore, we
get
\begin{description}
\item{\sf Dimension 0:} $c,\xi,\tphi, k^i, r^i.$
\item{\sf Dimension 1:} $A,\eta,\btpsi,\htpsi,\tpsi.$
\item{\sf Dimension 2:} $B,\bc,\bxi,h_c,h_\xi,\btphione,
\hone,\btphitwo,\htwo, \bk^i,h_k^i,\bro i,\hro i,\brt i,\hrt i.$
\end{description}

\paragraph{The explicit computation}
Our first task is to compute $s\Psi$. This will produce
many terms which we can divide into two classes:
terms that contain a Lagrange multiplier (the $h$-fields) and
terms that do not. The former will impose the gauge-fixing
conditions \rf{covetaB}, \rf{covB},
\rf{covtpsi} and \rf{covbtpsi}
(notice that---and this is the advantage of working in the
Landau gauge---we do not have quadratic terms in the $h$s, so
the $h$-integrations produce $\delta$-functionals of the
constraints). 
In the computation of the latter,
several terms will be canceled after explicitly imposing 
these gauge-fixing conditions. In particular, all the terms
containing the ghost $c$ (apart from those in $s\PsiYM$) are killed
since the covariant gauge-fixing conditions are $\calG$-equivariant;
e.g., in the $s$-variation of $\braket\bxi{d_A^*\eta}$, we will
remove the term $\braket\bxi{[d_A^*\eta,c]}$ by imposing
$d_A^*\eta=0$. Particular care has to be taken in the variation
of the last line in \rf{Psi} since $\btphitwo$ is not 
$\calG$-equivariant; the $c$-dependent part will then read
(by adding and subtracting $d_A[\btphitwo,c]$)
\[
\braket\btpsi{[d_A^*B+d_A\btphitwo+\brt i\omega_i[A],c]
-d_A[\btphitwo,c]}.
\]
The first term then vanishes by the gauge-fixing condition
\rf{covB} of $B$, while
the last term can be rewritten as $\braket{d_A^*\btpsi}{[\btphitwo,c]}$
and vanishes by the gauge-fixing conditions
\rf{covbtpsi} of $\btpsi$.

By imposing the gauge-fixing conditions,
we can also simplify the action $\SBFYM$: 
the effect is to eliminate the mixed term in $B$ and $\eta$.

Finally, we see that, thanks to the gauge-fixing conditions, we
can always replace $d_A^*d_A$ by the invertible
operator  $\Delta_A'$ defined as
\beq
\Delta_A' = 
\Delta_A + \pi_{{\rm Harm}_A} =
\left\{
\begin{array}{lcl}
1 &\quad& \mbox{on $\ker(\Delta_A)={\rm Harm}_A$}\\
\Delta_A &\quad& \mbox{on ${\rm coker}(\Delta_A)$}
\end{array}
\right.
\lab{DeltaA'}
\eeq
where $\pi_{{\rm Harm}_A}$ is the projection to harmonic
forms.
Notice that $\Delta_A'=\Delta_A$ on zero-forms since $A$ is an 
irreducible connection.
In the following, we will denote by $G_A$  the inverse of
$\Delta_A'$ and by $\det'(\Delta_A)$ the determinant of $\Delta_A'$.

Therefore, the gauge-fixed action---after all these 
simplifications---reads
\beq
\SBFYM^{\rm cov.\,g.f.} \begin{array}[t]{cl} 
= & i\braket B{*F_A} + \gym^2 \braket BB + 
\frac12\braket\eta{\Delta_A'\eta} + \\
+ & i\Big( s\PsiYM +
\braket{h_\xi}{d_A^*\eta} + h_k^i\braket{\omega_i[A]}\eta +\\
+ & \braket\hone{d_A^*\tpsi} 
+ \hro i\braket{\omega_i[A]}\tpsi +\\
+ & \braket\htpsi{d_A^*B+d_A\btphitwo+\brt i\,\omega_i[A]} + \\
+ & \braket\htwo{d_A^*\btpsi} 
+ \hrt i\braket{\omega_i[A]}\btpsi +\\
- & \braket\bxi{\Delta_A\xi} - \bk^i k^j \delta_{ij} \sqrt V +\\
+ & \braket\btphione{\Delta_A\tphi} + 
\bro i r^j \delta_{ij} \sqrt V +\\
- & \sqtg\braket{d_A\btpsi}{[F_A,\xi]}+
\braket\btpsi{\Delta_A'\tpsi}
\Big).
\end{array}
\lab{SBFYMgf}
\eeq
Notice that there is only one term which is singular as $\gym\to0$.
However, this singularity can be removed easily if one rescales
$\xi\to \gym\,\xi$ and $\bxi\to\bxi/\gym$. 

Now we can start integrating out fields in order to prove 
\rf{BFYMYM}. We want to point out that it is not necessary
to choose a background for $\eta$ and $B$ since they already
belong to vector spaces.

\subparagraph{Step 1} Integrate $\bk^i,k^i,\bro i,r^i$.

The integration over the first two variables yields $V^{\tb 1[A]/2}$,
while the integration over the last two of them yields
$V^{-\tb 1[A]/2}$; therefore, the contributions cancel each other.

\subparagraph{Step 2} Integrate $\bxi,\xi,\btphione,\tphi$.

The first integration yields $\det\Delta_A^{(0)}$, while
the second yields $(\det\Delta_A^{(0)})^{-1}$ and they cancel
each other. Notice that
there are no sources in $\bxi$, so the integration kills
the term in $\xi$.

\subparagraph{Step 3} Integrate $\btpsi,\tpsi, \hone, \htwo,
\hro i, \hrt i$.

The integration over the first two fields yields ${\det}'\Delta_A^{(1)}$.
Moreover, since there are linear sources in $\btpsi$ and $\tpsi$,
viz.,
\[
i\braket{d_A\hone+\hro i\omega_i[A]}\tpsi
-i\braket\btpsi{d_A\htwo+\hrt i\omega_i[A]},
\]
the Gaussian integration will give the following contribution
to the action
\beq
i\left({
\braket\hone{X_A\htwo} + \hro i\hrt j\delta_{ij}\sqrt V
}\right),
\eeq
where
\beq
X_A = d_A^* G_A d_A : \Omega^0(M, \ad P) \to \Omega^0(M, \ad P).
\lab{xia}
\eeq
Notice that there are no other terms since $G_A\omega_i[A]=
\omega_i[A]$ and $d_A^*\omega_i[A]=0$.

Now the integration over $\hone$ and $\htwo$ yields $\det X_A$
(we will show shortly that $X_A$ is invertible),
while the integration over $\hro i$ and $\hrt i$ yields
$V^{\tb 1[A]/2}$.

Therefore, the net contribution of this step is given by
\[{\det}'\Delta_A^{(1)}\, \det X_A\, V^{\tb 1[A]/2}.\]

\subparagraph{Step 4} Integrate $\eta, h_\xi, h_k^i$.

The first integration yields $({\det}'\Delta_A^{(1)})^{-1/2}$;
moreover, the linear source in $\eta$, viz.,
\[
i\braket{d_A h_\xi+h_k^i\omega_i[A]}\eta,
\]
produces the following contribution to the action:
\[
\braket{h_\xi}{X_A h_\xi} + h_k^i h_k^j\delta_{ij}\sqrt V.
\]
Then the $h_\xi$-integration yields $(\det X_A)^{-1/2}$, while
the $h_k^i$-integrations yield $(4V)^{-\tb 1[A]/4}$.

Therefore, the net contribution of this step is given by
\[({\det}'\Delta_A^{(1)}\,\det X_A)^{-1/2}\, (4V)^{-\tb 1[A]/4}.\]

\subparagraph{Step 5} Integrate $B$.

Apart from an irrelevant $\gym$-dependent factor, 
the Gaussian $B$-integration with source
\[
i\braket B{*F_A + d_A\htpsi}
\]
gives the following contribution to the action
\beq
\frac1{4\gym^2}\left(
\braket{F_A}{F_A}+\braket\htpsi{\Delta_A'\htpsi}
\right),
\lab{Fhtpsi}
\eeq
the mixed terms disappearing because of the Bianchi identity.

\subparagraph{Step 6} Integrate $\htpsi,\btphitwo,\brt i$.

The $\htpsi$-integration with quadratic term given in \rf{Fhtpsi}
and source
\[
i\braket\htpsi{d_A\btphitwo + \brt i\omega_i[A]}
\]
yields $({\det}'\Delta_A^{(1)})^{-1/2}$ plus the contribution
\[
\gym^2\left(
\braket\btphitwo{X_A\btphitwo} + \brt i\brt j\delta_{ij}\sqrt V
\right).
\]
Then the remaining integrations yield $(\det X_A)^{-1/2}$ and
$(4\gym4V)^{-\tb 1[A]/4}$.

Therefore, the net contribution of this step is 
\[({\det}'\Delta_A^{(1)})^{-1/2}\, 
(\det X_A)^{-1/2}\, (4\gym4V)^{-\tb 1[A]/4}.\]

\subparagraph{The operator $X_A$ is invertible}
In order to complete all the steps in the functional integration,
we still have to prove that the operator $X_A$ is invertible.

Let us represent the space $\Omega^1(M,\ad P)$ as the sum of three orthogonal 
subspaces: the vertical subspace 
$V_A=d_A\Omega^0(M,\ad P)$, ${\rm Harm}_A^1(M,\ad P)$,
and $\hat H_A \equiv H_A\ominus {\rm Harm}_A^1(M,\ad P)$ 
where $H_A=\ker d_A^*$ is the
horizontal subspace. Here horizontality and verticality are defined with
respect to the connection form on $\calA$ given by
\[G_A\,d_A^*:\Omega^1(M,\ad P)
\to \Omega^0(M,\ad P).\] 
The operator $\Delta^{(1)}_A:\hat H_A \oplus V_A \to \hat H_A \oplus V_A$
is injective and satisfies the following relation:
\[\braket{\eta_1}{\Delta^{(1)}_A \eta_2}=\braket{(d_A + d_A^*)\eta_1}
{(d_A + d_A^*)\eta_2}.\] 
Hence both $\Delta^{(1)}_A$ and its inverse $G_A$ are (formally)
self-adjoint and positive. We can consider the  (formal)
``square root" $G_A^{1/2}$ and have, for any $\zeta\in\Omega^0(M,\ad P)$,
\[ \braket{\zeta}{X_A\zeta}=
\braket{d_A\zeta}{G_Ad_A\zeta}=||G_A^{1/2}d_A\zeta||^2.\]
Hence $X_A$ is invertible.

\subparagraph{Conclusions}
Collecting all the contributions, we get YM as the effective action.
All the determinants cancel each other and the only net contribution
of all the integrations is a factor $(2\gym)^{-\tb 1[A]}$. 
However, $\tb 1[A]$ is constant in the neighborhood $\calN$. 
If this neighborhood does not coincide with $\calA$, one can choose
an interpolating gauge that smoothly connects the covariant gauge in
the interior of $\calN$ with the trivial gauge outside.
This ends the proof of 
\rf{BFYMYM} in the covariant gauge.

\subsubsection{The perturbative expansion}
As we have remarked after eqn.\ \rf{SBFYMgf}, a suitable rescaling
of $\xi$ and $\bxi$ removes any singularity as $\gym\to0$. This allows
weak-coupling perturbation theory.

To start with, one has to consider the fluctuations $\alpha$,
$e$ and $\beta$ of the fields $A,\eta,B$; viz.,
\beq
\begin{array}{lcl}
A &=& A_0 + q\alpha,\\
\eta&=& \eta_0 + e,\\
B &=& B_0 + \frac 1q\beta,
\end{array}
\lab{fluct}
\eeq
where $q$ is a free parameter, and
$(A_0,\eta_0,B_0)$ is a critical point of the action:
i.e., $A_0$ is a critical connection, $\eta_0=0$ and $B_0$
is given by \rf{covBsol}.
On the fluctuations, the covariant gauge fixing reads
\beq
\begin{array}{lcl}
d_{A_0}^* e + O(q) &=& 0, \quad e\perp{\rm Harm}_A^1(M, \ad P),\\
d_{A_0}^*\beta+q^2*\wcomm\alpha{*B_0} + O(q)
&\in& d_A\Omega^0(M, \ad P)
+ {\rm Harm}_A^1(M, \ad P).
\end{array}
\eeq
[Recall that $q^2B_0=O(1)$.]

\paragraph{The general case}
The quadratic part of the gauge-fixed action \rf{SBFYMgf} reads
\[
i\braket\beta{*d_{A_0}\alpha} + \frac12\left(\frac q\gym\right)^2
\braket{F_{A_0}}{\alpha\wedge\alpha} +
\left(\frac\gym q\right)^2\braket\beta\beta +
\frac12\braket e{\Delta_{A_0}'e} 
\]
plus the gauge-fixing terms.
Therefore, we see that the $\alpha\beta$- and $e$-sectors decouple.
Since we have both a term in $\gym/q$ and one in $q/\gym$, we must
take $q\sim\gym$ (a convenient choice is $q=\sqrt2\,\gym$). 

\paragraph{Perturbative expansion around a flat connection}
If the connection $A_0$ is flat, then $B_0=F_{A_0}=0$ and there
are no terms in $q/\gym$ in the quadratic part of the action.
Therefore, we can also take $\gym\ll q\ll 1$
and consider $(\gym/q)^2\braket\beta\beta$ as a perturbation.
More precisely, we take 
\beq
i\braket\beta{*d_{A_0}\alpha} +
\frac12\braket e{\Delta_{A_0}'e} + \mbox{gauge-fixing terms}
\lab{BFYM0q}
\eeq
as unperturbed action.
This is possible since the quadratic form in \rf{BFYM0q}
is non-degenerate.
In fact, the kernel is determined by the conditions
\[ 
d_{A_0}\beta=0,\quad d_{A_0}^*\beta + d_{A_0}\btphitwo = 0.
\]
(Notice that $b^1[A_0]=0$ if $A_0$ is flat.) 

Applying $d_{A_0}$ to the second equation we get $\Delta_{A_0}\beta=0$.
The kernel is therefore empty if there are no harmonic two forms
(and in general is finite dimensional). 

The propagators can be computed easily.
The $\alpha\beta$ propagator (i.e., the inverse of $*d_{A_0}$ on its
image) is the same as in pure $BF$ theory in the covariant gauge, 
as is clear by
comparing \rf{BFYM0q} with \rf{pBFq}; viz., it is 
the integral kernel of
(generalized) Gauss linking numbers.\footnote{Recall that 
in four dimensions we have
linking numbers between spheres and loops as opposed to the standard
linking numbers between loops in three 
dimensions}\ \ The $ee$-propagator is the same as the 
propagator for the
fluctuation of the connection in YM theory, as is clear by
comparing \rf{BFYM0q} with \rf{YMq}.

The perturbative expansion will then be organized as a 
formal double expansion
in $q$ and $(\gym/q)$. Notice that the theory is however independent of
$q$; in fact, a rescaling $q\to tq$ can be reabsorbed by the rescaling
$\alpha\to \alpha/t, \beta\to t\beta, \htpsi\to\htpsi/t$.
This reflects the analogous independency on the coupling constant
found in pure $BF$ theory in any dimension.


It is conceivable that quantization might break this symmetry
if we consider $B$-dependent observables. 
(The equivalence with YM theory rules out this possibility when we
consider only YM observables.)

\subsection{The self-dual gauge fixing}\lab{sec-sd}
\paragraph{Preliminaries}
The space of two-forms can canonically be split into the
sum of self-dual and anti-self-dual forms [denoted by $\Omplus$
and $\Omminus$] which satisfy $P^+ \omega = \omega$ and
$P^- \omega = \omega$ respectively, where the projection operators
$P^+$ and $P^-$ are defined as
\beq
P^\pm = \frac{1\pm*}2.
\eeq

By using one of these projection
operators (whatever follows is true by replacing
self-duality by anti-self-duality everywhere), we can define 
a new operator $D_A$ on the complex 
$\Omega^*(M,\ad P)\ominus\Omminus$ as
\beq
D_A := \left\{
\begin{array}{ccccc}
d_A &:& \Omega^0(M,\ad P) &\to& \Omega^1(M,\ad P)\\
\sqrt2 P^+ d_A &:& \Omega^1(M,\ad P) &\to& \Omplus\\
\sqrt2 d_A &:& \Omplus &\to& \Omega^3(M,\ad P)\\
d_A &:& \Omega^3(M,\ad P) &\to& \Omega^4(M,\ad P)
\end{array}
\right.
\lab{deTA}
\eeq
Then we can define the elliptic operator
\beq
\tDelta_A = D_A^*D_A + D_AD_A^*,
\eeq
and prove the following identities for this deformed
Laplace operator on forms of various degrees:
\beq
\begin{array}{lcl}
\tDelta_A^{(0)} &=& \Delta_A^{(0)},\\
\tDelta_A^{(1)} &=& \Delta_A^{(1)} - *\wcomm{F_A}{\ },\\
\tDelta_A^{(2)} &=& 2D_AD_A^* = 2D_A^*D_A = 2 P^+\Delta_A^{(2)}P^+.
\end{array}
\lab{tDelta}
\eeq
Since we are considering only irreducible
connections, the (deformed)
Laplace operator is invertible on zero forms.

We will denote by $\tHarm_A(M,\ad P)$ the (finite) kernel of 
$\tDelta_A$. Notice that
\beq
\begin{array}{lcl}
\tHarm_A^0(M,\ad P) &=& {\rm Harm}_A^0(M,\ad P)=\{0\},\\
\tHarm_A^1(M,\ad P) &\supset& {\rm Harm}_A^1(M,\ad P),\\
\tHarm_A^2(M,\ad P) &=& P^+{\rm Harm}_A^2(M,\ad P);
\end{array}
\eeq
As in the case of the ordinary covariant Laplacian, see \rf{DeltaA'},
we can define the invertible operator
\beq
\tDelta_A' = 
\tDelta_A + \pi_{\tHarm_A}=
\left\{
\begin{array}{lcl}
1 &\quad& \mbox{on $\ker(\tDelta_A)=\tHarm_A$}\\
\tDelta_A &\quad& \mbox{on ${\rm coker}(\tDelta_A)$}
\end{array}
\right.
\lab{tDeltaA'}
\eeq
and its inverse $\tG_A$.

Finally, if $A$ is a non-trivial anti-self-dual connection (i.e., $P^+F_A=0$),
then we assume that  
\[ D_A:\Omega^1(M,\ad P) \to \Omplus \]
is surjective, or equivalently, that 
\beq
\ker(D_A^*)=\{0\};\quad D_A^*:\Omplus\to \Omega^1(M,\ad P).
\lab{surject}
\eeq
Notice that \rf{surject} is verified for a
dense set of (conformal classes of) metrics for $G=SU(2)$ \cite{Donaldson}. 

For any $\sigma\in \Omega^1(M,\ad P)$ we have
\[D_{A+t \sigma}^*= D_A^* + t Q_{\sigma},\]
where the $A$-independent operator $Q_{\sigma}$ 
is defined by:
\[
Q_{\sigma}\varphi\equiv \sqrt{2}*[\sigma,\varphi]. \quad\varphi\in \Omplus.\]
This implies that
for $t$ sufficiently small $D^*_{A+t\sigma}$ is also invertible  and
that
there is neighborhood of the space $\calM^-$ of anti-self-dual 
connections---which we will denote by $\calN$---where
the  property \rf{surject} holds. We take $\calN$ to be the inverse image
of a neighborhood of the moduli space.
By \rf{surject} it follows that $\tDelta_A^{(2)}$ is
invertible if $A\in\calN$, so $\tHarm_A^2(M,\ad P)=\{0\}$. 

If $A$ is an anti-self-dual connection, 
then $\tHarm_A^1(M,\ad P)=T_A\calM^-$; 
therefore, for $A$ in a neighborhood of $\calM^-$, 
$\dim\tHarm_A^1(M,\ad P)=\dim\calM^-=m^-$. Moreover, $D_A^2=F_A^+=0:
\Omega^0(M,\ad P)\to\Omplus$. This implies that
\[
({\rm Im}\,D_A^*\bigcap{\rm coker}\,D_A^*)\bigcap\Omega^1(M,\ad P)=\{0\}
\quad\mbox{if $A\in\calM^-$}.
\]
Therefore, \rf{surject} is an injection from $\Omplus$ to
${\rm ker}\,D_A^*\bigcap\Omega^1(M,\ad P)$. 
By continuity this property will hold
in a neighborhood of $\calM^-$. We will denote by $\calN'$ the intersection
of this neighborhood with $\calN$ and with the neighborhood where
$\dim\tHarm_A^1(M,\ad P)$ is constant. 

Therefore, the neighborhood
$\calN'$---which we will use in the rest of this section---is
characterized by the following two properties:
\begin{enumerate}
\item $D_A : ({\rm ker}\,D_A^*\bigcap\Omega^1(M,\ad P))
\to\Omplus$ is surjective
if $A\in\calN'$;
\item $\dim\tHarm_A^1(M,\ad P)=m^-$ if $A\in\calN'$.
\end{enumerate}

\paragraph{The definition of the self-dual gauge fixing}
Now we are in a position to define the self-dual gauge fixing 
(for further details, s.\ subsec.\ \ref{subsec-gf})
in terms
of a gauge-fixing condition on the connection $A\in\calN'$ 
together with the conditions
\beq
D_A^*\eta=0,\quad \eta\perp\tHarm^1_A(M, \ad P),\quad
P^+B=0,
\lab{sdetaB}
\eeq
and, by consistency,
\beq
D_A^*\tau=0,\quad \tau\perp\tHarm^1_A(M, \ad P).
\lab{sdtau}
\eeq
In the context of BRST quantization, the last conditions will imply
\beq
D_A^*\tpsi=0,\quad \tpsi\perp\tHarm^1_A(M, \ad P).
\lab{sdtpsi}
\eeq

Also in this case we have gauge fixings 
which  are interpolating between
\rf{sdetaB} and the trivial gauge fixing $\eta=0.$
In fact, the trivial gauge fixing can be written  as
\[
D_A^*\eta=0,\quad \eta\perp\tHarm^1_A(M, \ad P),\quad
D_A\eta=0.
\]
The interpolating gauge fixings can be then written as 
\[
\lambda P^+B + (1-\lambda) D_A\eta = 0,
\]
with $\lambda\in [0,1]$.

Again one might also choose $\lambda$ to be smooth but not constant 
on $\calA$. 
In particular, if we choose $\lambda$ to be constant and equal to 1
in an open neighborhood of $\calM^-$ contained
in the neighborhood $\calN'$, and constant and equal to 0 outside 
$\calN'$, we obtain a gauge fixing that is defined on the whole space 
$\calA$ and restricts to the self-dual gauge fixing close to the  
anti-self-dual connections.

\subsubsection{Classical equivalence}
First we observe that an anti-self-dual connection solves
the YM equations of motion. Then we see that the self-dual
part of the first equation of \rf{eom} reads
\[
D_A\eta=0,
\]
which together with the gauge-fixing conditions implies
\beq
\eta=0;
\lab{sdetasol}
\eeq
therefore, we get
\beq
B = -\frac i{2\gym^2} *F_A.
\lab{sdBsol}
\eeq
Notice that this solution is the same as those obtained
with the trivial and the covariant gauge fixings.

\subsubsection{Quantum equivalence}
To implement the self-dual gauge fixing, we have to introduce
a BRST complex which is slightly different from that
used for the covariant case. 

More precisely, we have to replace the pairs $(\btpsi, \btphitwo)$
and $(h_\tpsi,\htwo)$ respectively
by the self-dual antighost $\bchi^+$ (with ghost number $-1$)
and by the self-dual Lagrange multiplier $h_\chi^+$ (with
ghost number $0$). Notice that the number of degrees of freedom is
preserved; in fact, $\btpsi$ is a one-form with ghost number $-1$
(so four fermionic degrees of freedom), while $\btphitwo$
is a zero-form with ghost number $0$ (so one bosonic 
or, equivalently, minus one fermionic degree of freedom); this
gives three fermionic degrees of freedom which is consistent
with the fact that $\bchi^+$ is a self-dual two-form with
ghost number $-1$. A similar counting holds for the other
fields.

The BRST transformation rules for the antighosts and the Lagrange
multipliers are the same as those described in the case of the
covariant gauge fixing; as for the new fields, we have
\beq
s\bchi^+=h_\chi^+,\quad sh_\chi^+=0.
\eeq

To deal with the harmonic one-forms of the deformed Laplace
operator, we introduce an orthogonal
basis for $\tHarm_A^1(M,\ad P)$---which
we still denote by $\omega_i[A]$, $i=1,\ldots,m^-$---and normalize it
as in \rf{omnorm}.

As in the case of the covariant gauge fixing, we introduce new constant
ghosts $k^i$ and $r^i$, together with their antighosts and
Lagrange multipliers, with BRST transformation rules given by
\rf{skr} and \rf{skrb}. Moreover, we
rewrite the BRST transformations
for $\eta$ and $\tpsi$ as in \rf{etatpsih}.

The self-dual gauge fixing is eventually implemented by
choosing the following gauge-fixing fermion:
\beq
\Psi \begin{array}[t]{cl} 
= & \PsiYM + \\
+ & \braket\bxi{D_A^*\eta} + \bk^i\braket{\omega_i[A]}\eta +\\
+ & \braket\btphione{D_A^*\tpsi} 
+ \bro i\braket{\omega_i[A]}\tpsi +\\
+ & \braket{\bchi^+}B,
\end{array}
\lab{sdPsi}
\eeq

The canonical dimensions of the old fields are the same as in the
case of the covariant gauge fixing, while the new fields $\bchi^+$ and
$h_\chi^+$ have both dimension two.

\paragraph{The explicit computation}
As in the computation with the covariant gauge fixing, the gauge-fixed
action $\SBFYM+is\Psi$ can be simplified if one imposes the
gauge-fixing conditions explicitly. At the end  we get
\beq
\SBFYM^{\rm s.d.\,g.f.} \begin{array}[t]{cl} 
= & -i\braket {B^-}{P^- F_A} + \gym^2 \braket {B^-}{B^-} +\\ 
+&\frac12\braket\eta{\Delta_A\eta} - 
\sqrt2\gym \braket{B^-}{P^-(d_A\eta)} +
\\
+ & i\Big( s\PsiYM +
\braket{h_\xi}{D_A^*\eta} + h_k^i\braket{\omega_i[A]}\eta +\\
+ & \braket\hone{D_A^*\tpsi} 
+ \hro i\braket{\omega_i[A]}\tpsi 
+  \braket{h_\chi^+}{B^+} + \\
- & \braket\bxi{\Delta_A\xi} - \bk^i k^j \delta_{ij} \sqrt V +\\
+ & \braket\btphione{\Delta_A\tphi} + 
\bro i r^j \delta_{ij} \sqrt V +\\
- & \sqtg\braket{\bchi^+}{P^+[F_A,\xi]}+
\braket{\bchi^+}{D_A\tpsi}
\Big),
\end{array}
\lab{SBFYMgfsd}
\eeq
where $B^\pm$ are the self-dual and anti-self-dual components of $B$.

Notice that there is only one term which is singular as $\gym\to0$.
However, this singularity can be easily removed if one rescales
$\xi\to \gym\,\xi$ and $\bxi\to\bxi/\gym$. 

Now we can start integrating out the fields.

\subparagraph{Step 1} Integrate $\bk^i,k^i,\bro i,r^i$.

As in the case of the covariant gauge fixing, this integration
gives no contribution.

\subparagraph{Step 2} Integrate $\bxi,\xi,\btphione,\tphi$.

Again, this integration does not contribute.

\subparagraph{Step 3} Integrate $\bchi,\tpsi,\hone, \hro i$.

The relevant terms in the action can be written as
\[
\frac i2 \braket{X}{MX},
\]
where $X$ is the vector
\[
X = \left(\begin{array}{c}
\tpsi\\
{\bf h}_{r_1}\\
\bchi^+\\
\hone
\end{array}\right)
\in\Omega^1\oplus{\bf R}^{m^-}\oplus\Omega^{2,+}\oplus\Omega^0,
\]
and $M$ is the anti-hermitean operator
\beq
M = \left(\begin{array}{cccc}
0 & -\bomega_A & -D_A^* & -D_A\\
\bomega_A^* & 0 & 0 & 0\\
D_A & 0 & 0 & 0\\
D_A^* & 0 & 0 & 0
\end{array}\right).
\lab{defM}
\eeq
The scalar product is defined as in \rf{scalar} on $\Omega^*(M,\ad P)$
and is the ordinary Euclidean scalar product on ${\bf R}^{m^-}$.

The operator $\bomega_A : {\bf R}^{m^-}\to \Omega^1(M,\ad P)$ is
defined by
\[
\bomega_A{\bf h}_{r_1} = \sum_{i=1}^{m^-}
\omega_i[A]\,\hro i,
\]
and its adjoint $\bomega_A^*:\Omega^1(M,\ad P)\to{\bf R}^{m^-}$
acts as
\[
(\bomega_A^*\tpsi)_i = \braket{\omega_i[A]}\tpsi.
\]

The functional integration will then produce the Pfaffian of $M$
which, as we will prove in App.\ \ref{app-Pf}, is given, up
to an irrelevant constant, by
\beq
{\rm Pf}(M) \propto
(\det(\Delta_A^{(0)}-R_A)\,
{\det}' \tDelta_A^{(1)}\,\det\Delta_A^{(2,+)})^{1/4}\,
\,V^{m^-/8},
\lab{Pf}
\eeq
where
\beq
R_A = D_A^*\pi_{{\rm coker}(D_A)}D_A : 
\Omega^0(M,\ad P)\to\Omega^0(M,\ad P).
\lab{defR}
\eeq
Notice that the operator
\beq
\widehat\Delta_A = \Delta_A - R_A : 
\Omega^0(M,\ad P)\to\Omega^0(M,\ad P)
\lab{defhDelta}
\eeq
is invertible for $A\in\calN'$ (s.\ App.\ \ref{sec-pq}).

\subparagraph{Step 4} Integrate $h_\chi^+,B$.

First notice that, since self-dual and anti-self-dual two-forms
are orthogonal, the integration over $B$ can be replaced by an
integration over $B^+$ and $B^-$ with Jacobian equal to 1.

The $(h_\chi^+,B^+)$-integration is then trivial. The $B^-$-integration
with source
\[
\braket{B^-}{P^-(-iF_A-\sqrt2\gym\,d_A\eta
)}
\]
yields a constant term depending on $\gym$ (of which we do not care)
plus the following contribution to the action
\[
\frac1{4\gym^2}\,\braket{P^-F_A}{P^-F_A} -
\frac i{\sqrt2\,\gym}\,\braket{P^-F_A}{P^-(d_A\eta)} -
\frac12 \braket{P^-(d_A\eta)}{P^-(d_A\eta)}.
\]

Therefore, at this stage we get the following effective action:
\[
\frac1{4\gym^2}\,\braket{P^-F_A}{P^-F_A} +
\frac14\braket\eta{\tDelta_A'\eta} +
i\braket\eta{\frac{-1}{2\gym}\,D_A^*P^+F_A
+D_Ah_\xi+h_k^i\omega_i[A]} +is\PsiYM,
\]
where we have used the fact that 
$\sqrt2d_A^*P^-F_A=\sqrt2d_A^*P^+F_A=D_A^*P^+F_A$.

\subparagraph{Step 5} $\eta,h_\xi,h_k^i$.

The $\eta$-integration yields $({\det}'\tDelta_A^{(1)})^{-1/2}$ plus
the following contribution to the action:
\[
\frac1{4\gym^2}\,\braket{P^+F_A}{\tZ_AP^+F_A}-
\frac1\gym\,\braket{h_\xi}{D_A^*\tG_AD_A^*P^+F_A} +
\braket{h_\xi}{\tX_A h_\xi}
+h_k^ih_k^j\delta_{ij}\sqrt V,
\]
where
\beq
\begin{array}{ccccccc}
\tX_A&=&D_A^*\tG_AD_A &:& \Omega^0(M,\ad P)&\to&\Omega^0(M,\ad P),\\
\tZ_A&=&D_A\tG_AD_A^* &:& \Omplus&\to&\Omplus.
\end{array}
\lab{deftXZ}
\eeq
Even though $D_A$ does not commute with $G_A$ (unless $A\in\calM^-$),
these two operators are identity operators as long as $A\in\calN'$.
For details, s.\ App.\ \ref{sec-pq}.

The $h_k^i$-integrations yield $(4V)^{-m^-/4}$, while
the $h_\xi$-integration produces $(\det\tX_A)^{-1/2}=1$ plus
the contribution
\[
-\frac1{4\gym^2}\,
\braket{P^+F_A}{\widehat Z_AP^+F_A},
\]
where
\beq
\widehat Z_A = D_A\tG_AD_AD_A^*\tG_AD_A^* :
\Omplus\to\Omplus.
\lab{defhZ}
\eeq
As long as $A\in\calN'$, this operator is null as proved in 
App.\ \ref{sec-pq}.

\subparagraph{Conclusions} 
Putting together the determinants
coming from steps 3 and 5 we find a net contribution
\beq
J[A] = \frac{(\det(\Delta_A^{(0)}-R_A))^{1/4}
\,(\det\Delta_A^{(2,+)})^{1/4}}
{({\det}'\tDelta_A^{(1)})^{1/4}}.
\lab{defJ}
\eeq
In App.\ \ref{sec-pq}, we show that $J[A]=1$ if $A\in\calN'$.
Moreover, Step 4 and Step 5 reconstruct YM action in the form
\[
\SYM[A] = \frac1{4\gym^2}\,\braket{P^-F_A}{P^-F_A} +
\frac1{4\gym^2}\,\braket{P^+F_A}{P^+F_A}. 
\]
Therefore, we have proved the equivalence between
BFYM and YM theory (for $A\in\calN'$)
by using the self-dual gauge fixing. More explicitly, we have
shown that
\beq
\int_{(T\calN'\times T_{F_A}\calB)/\Gaff,{\rm self-dual}}
\exp(-\SBFYM[A,\eta,B])\ \calO[A]\propto
\int_{\calN'/\calG} \exp(-\SYM[A])\ \calO[A].
\lab{BFYMYMsd}
\eeq

If we choose a gauge fixing that restricts to
the self-dual gauge in the interior of $\calN'$ and to the trivial gauge
outside, we can extend the equivalence to the whole $\calA$.

\subsubsection{The perturbative expansion around an anti-self-dual
connection}
Again we can consider fluctuations around a background
as in \rf{fluct}. Since we assume the connection not to be flat,
we will have both terms in $q/\gym$ and in $\gym/q$, so we must
take $q\sim\gym$. It is convenient to choose $q=\sqrt2\,\gym$.

The gauge-fixing conditions \rf{sdetaB} on the fluctuations
simply read
\[
d_{A_0}^*\eta + O(\gym) = 0, \quad
\eta \perp\tHarm_{A_0}^1(M,\ad P),\quad
\beta^+=0.
\]
The quadratic part of the gauge-fixed action \rf{SBFYMgfsd}
reads
\[
\begin{array}{l}
-i\braket{\beta^-}{d_{A_0}\alpha} +
\braket{F_{A_0}}{\alpha\wedge\alpha} +
\frac12\braket{\beta^-}{\beta^-} +\\
-\braket{\beta^-}{d_{A_0}e} - 
i\braket{F_{A_0}}{[\alpha,e]} +
\frac12\braket e{\tDelta_{A_0}'e + *[F_{A_0},e]}
,
\end{array}
\]
plus the gauge-fixing terms. Unlike in the case of the covariant
gauge fixing, the $\alpha\beta$- and $e$-sectors do not
decouple. However, if we perform the change of variables
\beq
\alpha' = \alpha -ie,\quad
e' = e, \quad
\beta' = \beta,
\eeq
the quadratic part of the action turns out to be
\beq
-i\braket{(\beta')^-}{d_{A_0}\alpha'} +
\braket{F_{A_0}}{\alpha'\wedge\alpha'} +
\frac12\braket{(\beta')^-}{(\beta')^-} +
\frac12\braket {e'}{\tDelta_{A_0}'e'},
\lab{BFYMsdq}
\eeq
plus the gauge-fixing terms. Now the 
$\alpha'\beta'$- and $e'$-sectors decouple. Moreover,
for the $\alpha'\beta'$-sector we recognize the propagators
of the topological $BF$ theory with a cosmological term
in the self-dual gauge, see. \rf{BFtcq},
whereas the $e'e'$-propagator turns out to be same as the propagator
for the fluctuation of the connection in YM theory [thanks to
\rf{cDelta} and to the second equation of \rf{tDelta}].

\section{The relation with the topological $BF$ theories}
\lab{sec-BV}
In the previous section, studying
perturbative BFYM theory in the covariant gauge around 
a flat connection or in the self-dual gauge around a non-trivial
anti-self-dual connection, we have discovered
that a sector of the theory
corresponds to the topological $BF$ theory (pure or, respectively,
with a cosmological term) in the same gauge.

In this section we will recall the properties of the topological
$BF$ theories. The main problem with these theories is that
the symmetries are described by a BRST operator that is nilpotent
only on-shell. Therefore, one has to resort to the BV formalism
which we briefly introduce in subsection \ref{ssec-BV}.

We also want to discuss the relations between the BFYM and
the $BF$ theories before starting perturbation theory.
In the case of the self-dual gauge, this relation 
simply relies on the fact that $\braket BB = -\braket B{*B}$
when $B$ is anti-self-dual.

The case of the covariant gauge fixing with a flat background
connection is however more intricate, for it is related to
the limit $\gym\to0$ which is ill-defined as discussed
at the end of Sec.\ \ref{sec-foYM}.
We have already observed that in this limit 
the BFYM theory formally reduces
to the topological pure $BF$ theory plus a dynamical term for $\eta$,
s.\ eqn.\ \rf{SBFYM0}.

We have also observed that this limit is well-defined after 
fixing the gauge.

However, we would like this limit to be meaningful for the theory
even before a gauge is chosen. 

Dealing with the theory defined by \rf{SBFYM0} presents some 
difficulties. In fact, the term $\braket{d_A\eta}{d_A\eta}$
has a different symmetry on shell (the $T\calG$ action) and off shell
(only the $\calG$ action). Of course, one has to consider the larger
symmetry if one wants to quantize the theory. The on-shell symmetry
for $\eta$ can be made into an off-shell symmetry of the whole theory
by setting
\[
sB = [B,c] -d_A\tpsi + *[d_A\eta,\xi].
\]
However, now the BRST operator is nilpotent only on shell. (Notice that
this is a problem affecting pure $BF$ theory as well.)

Another way of seeing the problem is to perform the limit
$g^2\to0$ in BFYM theory. We meet the following difficulties:
\begin{enumerate}
\item There is no way of getting in the limit the previous
BRST transformation on $B$.
\item If we consider the BRST transformations as in \rf{BRSTBFYM},
we get, in the limit, the correct on-shell symmetry for $\eta$ but
a divergent transformation for $B$.
\item If we try to avoid this problem by rescaling 
$\xi\to \sqrt2\,\gym\,\xi$, we
get a well-defined transformation for $B$, but the transformation for
$\eta$ is correct only off shell now; this leads to contradictions when
we try to quantize the theory. In fact, if we decide not to fix the
gauge for $\eta$ we get in trouble when the curvature vanishes; 
on the other hand, if we want to gauge fix it, we have to introduce
the antighost $\bxi$, but then we get in trouble since the
$\bxi$-dependent terms in the gauge-fixed action are killed. (Of course,
if we first fix the gauge and then let $\gym\to0$, we do not have any
problem.) 
\item If we also decide to rescale $\eta\to\sqrt2\,\gym\,\eta$, the
quadratic term in $\eta$ disappears from the action. This means
that the symmetry on $\eta$ is given, as we correctly obtain,
by the whole $\Gaff$ action. However, now $B$ can be shifted by
$d_A\tpsi'$ with no relation between $\tpsi$ and $\tpsi'$. That is,
we have to introduce new ghosts.
\end{enumerate}

The solution to these problems is again
in the use of the BV formalism. 

\subsection{The BV formalism}
\lab{ssec-BV}
In the BV formalism, one considers the $\bf Z$-graded algebra of 
polynomials
in the {\em fields} $\{\Phi_i\}$ of the theory. We will denote by
$\epsilon(K)$ the grading---i.e., the {\em ghost number}---of 
the monomial $K$. As a shorthand notation, we will simply write 
$\epsilon_i$ for the ghost number of the generator $\Phi^i$.
Moreover, each field is given a Grassmann parity by the reduction
mod 2 of the ghost number (if half-integer-spin particles are present,
then their Grassmann parity is increased by one).

To each field $\Phi^i$ is then associated 
an antifield $\Phi_i^\dg$ which is completely equivalent to $\Phi^i$
under all respects but the ghost number; i.e., $\Phi_i^\dg$ is a section
of the same principal or associated bundle as $\Phi^i$ and is given
ghost number by
\beq
\epsilon(\Phi_i^\dg) = -\epsilon_i - 1.
\lab{ghn}
\eeq

\subsubsection{BV antibracket and Laplacian}
Given two functions $X$ and $Y$ of the variables
 $\{\Phi^i,\Phi_i^\dg\}$, one
defines the BV antibracket as
\beq
\antib XY := X \braket{\dphidb i}{\dphi i} Y
-X \braket{\dphib i}{\dphid i} Y
\lab{defantib}
\eeq
and the BV Laplacian as
\beq
\Delta X = \sum_i (-1)^{\epsilon_i} \braket{\dphid i}{\dphi i} X.
\lab{deflap}
\eeq
Notice that both the antibracket and the Laplacian increase the
ghost number by one. 

We remark that the two previous operations are not independent:
in fact, the BV antibraket can be written in terms of the BV
Laplacian and of the pointwise product of functions.

\subsubsection{Canonical transformations}
The BV formalism is defined modulo {\em canonical transformations},
i.e., transformations of the fields and antifields that 
preserve the BV Laplacian and, consequently, the BV antibracket.

A canonical transformation can
be obtained by introducing a generating functional 
$F(\Phi^i, \widetilde{\Phi^\dg_i})$, with $\epsilon(F) = -1$, 
such that
\beq
\widetilde{\Phi^i} = \frac{\overrightarrow\delta}
{\delta\widetilde{\Phi^\dg_i}}\,F,\quad
\Phi^\dg_i = \frac{\overrightarrow\delta}{\delta\Phi^i}\,F.
\lab{genfunc}
\eeq
In the BV context, there is no analogue of Liouville's theorem
in classical mechanics, and, in general,
the volume form is {\em not} preserved
by canonical transformations.  

Notice that rescalings of the form
$\Phi^i\to\lambda^i\Phi^i$, $\Phi_i^\dg=\Phi_i^\dg/\lambda^i$
are canonical transformations, their generating functional being
$F=\sum_i \lambda^i\braket{\Phi^i}{\widetilde{\Phi_i^\dg}}$.

\subsubsection{The implementation of symmetries}
Suppose we have an action $S[\varphi]$, where by $\varphi$ we denote
the classical fields (i.e., the zero-ghost-number fields that appear
in the action). The study of the on-shell symmetries allows the 
construction of the BRST complex (i.e., the whole set of fields 
$\Phi^i$) together with the BRST operator $s$. In many cases, this
operator turns out to be nilpotent also off shell, and the BRST 
formalism is enough to quantize the theory. However, there
are situations (e.g., in the $BF$ theories) where this is not true.
In these cases, the BV formalism provides a useful generalization
of the BRST formalism.

First of all one has to look for the BV action, i.e., a functional
$\SBV[\Phi,\Phi^\dg]$ that solves the {\em master equation}
\beq
\antib \SBV\SBV = 0,
\lab{me}
\eeq
and reduces to the classical action $S$ when the antifields are turned 
off, viz.,
\beq
\SBV[\Phi,0]=S[\varphi].
\lab{BVbc}
\eeq
In particular, one looks for a {\em proper solution} of \rf{me};
i.e., one requires the Hessian of $\SBV$ evaluated on-shell to have
rank equal to the number of fields. 

There is a theorem that states that, under some mild assumptions on $S$,
there exists one and only one (up to canonical transformations)
proper solution $\SBV$ to the master equation \rf{me} with boundary 
conditions \rf{BVbc}.

Thanks to the master equation and to
the properties of the BV antibracket, the operator
$\sigma$ defined by
\beq
\sigma X := (\SBV,X)
\eeq
turns out to be nilpotent. The boundary condition \rf{BVbc} then
ensures that, up to possible terms in the antifields,
$\sigma$ acts on the fields as $s$. 

If the BRST operator $s$ is nilpotent also off shell,
one can write the BV action as
\beq
\SBV[\Phi,\Phi^\dg] = S[\varphi] + 
\sum_i\braket{s\Phi^i}{\Phi^\dg_i}.
\lab{SBVBRST}
\eeq
In this case $\sigma=s$ on all fields.

\subsubsection{The BV quantization}
The quantization of the theory then proceeds by fixing the 
gauge.
This is achieved, as in the BRST formalism, by introducing a 
gauge-fixing fermion $\Psi[\Phi]$. Now, however, the gauge-fixed
action is defined by
\beq
S^{\rm g.f.}[\Phi] = \left.
\SBV[\Phi, \Phi^\dg]\right|_{\Phi^\dg_j = 
i\frac\delta{\delta \Phi^j}\Psi}.
\lab{BVgf}
\eeq
If $\SBV$ has the form \rf{SBVBRST}, then this procedure gives
$S^{\rm g.f.} = S + is\Psi$, as in the BRST formalism.

The condition that $\SBV$ should be
a proper solution of the master equation
makes perturbative quantization possible and---if 
$\Delta S=0$---independent of small deformations of 
$\Psi$.\footnote{
Usually one can choose $\Delta S=0$. However, the quantum corrections
due to renormalization generally break this condition.

To deal with this case, one has to consider the quantum BV action
$\SBV_\hbar(\Phi,\Phi^\dg)$ which satisfies the quantum master
equation
\[
\antib {\SBV_\hbar}{\SBV _\hbar} +2\hbar\Delta\SBV_\hbar = 0
\]
and the boundary condition
\[
\SBV_0(\Phi,\Phi^\dg) = \SBV(\Phi,\Phi^\dg).
\]
Notice that to a BV action there might correspond no quantum BV
action; in this case the theory is said to be anomalous.

When the theory is not anomalous,
fixing the gauge as in \rf{BVgf} with $\SBV$ replaced by $\SBV_\hbar$
yields a theory whose perturbative quantization is independent of
small deformations of $\Psi$.

Moreover, the quantum master equation implies that the operator
$\sigma_\hbar$ defined as
\[
\sigma_\hbar = \sigma +\hbar\Delta
\]
is nilpotent. Then one can show that the vacuum expectation value
of a functional $\calO_\hbar(\Phi,\Phi^\dg)$ such that 
$\sigma_\hbar\calO_\hbar=0$ is 
independent of small deformations of $\Psi$.
}
Moreover, the vacuum expectation value of a functional 
$\calO(\Phi,\Phi^\dg)$ such that $\sigma\calO=0$ turns out to be 
independent of small deformations of $\Psi$ as well.

\subsection{Applications of the BV formalism}
The theories we have considered in this paper---viz., first-
and second-order YM theory and BFYM theory---have a BRST operator
that closes also off shell; therefore, up to canonical transformations,
they can be written as in \rf{SBVBRST}. Explicitly, we have
\beq
\begin{array}{lcl}
\SYM^{\rm BV} &=& \SYM + \braket{d_Ac}{A^\dg} 
-\braket{{1\over 2}[c,c]}{c^\dg} + \braket{h_c}{\bc^\dg};\\
\SYMfo^{\rm BV} &=& \SYMfo + \braket{d_Ac}{A^\dg} + 
\braket{[E,c]}{E^\dg}
-\braket{{1\over 2}[c,c]}{c^\dg} + \braket{h_c}{\bc^\dg};\\
\SBFYM^{\rm BV} &=& \SBFYM + \braket{d_Ac}{A^\dg} +
\braket{[\eta,c] + d_A\xi - \sqrt2\, \gym\,\tpsi}{\eta^\dg} +\\
&+&\braket{[B,c] + \sqtg [F_A,\xi] - d_A\tpsi}{B^\dg} +\\
&+&
\braket{-{1\over 2}[c,c]}{c^\dg} +
\braket{-[\xi,c]+\sqrt2\, \gym\,\tphi}{\xi^\dg} +\\
&+& \braket{-[\tpsi,c]+d_A\tphi}{\tpsi^\dg}+
\braket{[\tphi,c]}{\tphi^\dg} + 
\sum_i\braket{h^i}{\bc_i^\dg},
\end{array}
\eeq
where, in the last line, we have denoted by $h^i$ and $\bc^i$ the
Lagrange multipliers and antighosts. 

The canonical transformation
$\xi\to\sqrt2\,\gym\,\xi,\xi^\dg\to\xi^\dg/(\sqrt2\, \gym)$ 
seems to remove
all the singularities from $\SBFYM^{\rm BV}$. However, in the $\gym\to0$
limit the BV action turns out not to be proper.
As a consequence, if we fix the gauge 
with $\Psi$ as in \rf{Psi}, we do not get the kinetic term for
$\bxi,\xi$ and quantization becomes impossible.

\subsubsection{The pure $BF$ theory}
The pure $BF$ theory is described by the action
\beq
S_{BF} = i\braket B{*F_A},
\eeq
and its symmetries are encoded by the following BRST transformations:
\beq
\begin{array}{lclclcl}
sA &=& d_Ac, &\quad& sc &=& -\frac12[c,c],\\
sB &=& [B,c] - d_A\tpsi, & & & & \\
s\tpsi&=&-[\tpsi,c]+d_A\tphi, &\quad& 
s\tphi &=& [\tphi,c].
\end{array}
\lab{BRSTBF}
\eeq
Notice that $s^2\not=0$ off shell; as a matter of fact, $s^2$ vanishes
on all fields but on $B$ where one gets
\[
s^2 B = -[F_A,\tphi].
\]
The BV action can be written as
\beq
\begin{array}{lcl}
S_{BF}^{\rm BV} 
&=& S_{BF}+ 
\braket{d_Ac}{A^\dg}
+\braket{[B,c] - d_A\tpsi -\frac12*[B^\dg,\tphi]}{B^\dg} +\\
&+&\braket{-{1\over 2}[c,c]}{c^\dg} +
\braket{-[\tpsi,c]+d_A\tphi}{\tpsi^\dg}+
\braket{[\tphi,c]}{\tphi^\dg} + 
\sum_i\braket{h^i}{\bc_i^\dg},
\end{array}
\lab{BFBV}
\eeq
where the sum is over the same antighosts and Lagrange multipliers
as in BFYM but $\bxi$ and $h_\xi$. It can be shown that \rf{BFBV}
is a proper solution of the master equation.

Notice that the BV action is not linear in $B^\dg$. This implies that
the operator $\sigma$ on $B$ acts in a different way than the operator
$s$; viz.,
\beq
\sigma B = [B,c] - d_A\tpsi - *[B^\dg,\tphi].
\lab{sigmaBBF}
\eeq
On all other fields, $\sigma=s$. Notice that $\sigma^2=0$ since
\[
\sigma B^\dg = -[B^\dg,c] - * F_A.
\]

\paragraph{Quantization in the covariant gauge}
The covariant gauge fixing for pure $BF$ theory is defined
exactly as in BFYM theory \rf{covetaB} and is quantistically
implemented by the same gauge-fixing fermion \rf{Psi}
(of course, forgetting of the conditions on $\eta$) using
\rf{BVgf}. After some algebra we can write the gauge-fixed
action as
\beq
S_{BF}^{\rm cov.\,g.f.} \begin{array}[t]{cl} 
= & i\braket B{*F_A} +  
i\Big( s\PsiYM 
+\\
+ & \braket\hone{d_A^*\tpsi} 
+ \hro i\braket{\omega_i[A]}\tpsi +\\
+ & \braket\htpsi{d_A^*B+d_A\btphitwo+\brt i\,\omega_i[A]} + \\
+ & \braket\htwo{d_A^*\btpsi} 
+ \hrt i\braket{\omega_i[A]}\btpsi +\\
+ & \braket\btphione{\Delta_A\tphi} + 
\bro i r^j \delta_{ij} \sqrt V +\\
+ & \frac12\braket{d_A\btpsi}{*[d_A\btpsi,\tphi]}+
\braket\btpsi{\Delta_A'\tpsi}
\Big).
\end{array}
\lab{SBFgf}
\eeq
As is usual in theories whose BRST operator is nilpotent
only on shell, a cubic term appears
in the ghost--antighost variables,
viz., $\braket{d_A\btpsi}{*[d_A\btpsi,\tphi]}$. This term is however
killed by the $\btphione\tphi$ integration since there are no sources
in $\btphione$. Recall that also in \rf{SBFYMgf} the term
$\braket{d_A\tpsi}{[F_A,\xi]}$ was irrelevant since the
$\bxi\xi$ integration killed it.
Therefore,
the gauge-fixing terms in \rf{SBFYMgf} are the same
as those which appears in \rf{SBFgf} plus the terms related
to the gauge fixing on $\eta$. Explicitly, we have
\beq
\SBFYM^{\rm cov.\,g.f.} = S_{BF}^{\rm cov.\,g.f.} + \gym^2\braket BB +
\frac12\braket\eta{\Delta_A'\eta} + 
is\left(\braket\bxi{d_A^*\eta} + \bk^i \braket{\omega_i[A]}\eta\right).
\lab{BFYM-BFgf}
\eeq 

\paragraph{Perturbative expansion}
The equations of motion of $BF$ theory are $F_A=d_AB=0$. 
Therefore, $A$ is a flat connection. In the covariant background, we 
also have $B=0$ (if there are harmonic two-forms, we have to require $B$
to be orthogonal to them).

If we then consider fluctuations 
around a critical background (i.e., $F_{A_0}=B_0=0$) as in \rf{fluct},
we get the quadratic action
\beq
i\braket{\beta}{*d_{A_0}\alpha} + \mbox{gauge-fixing terms},
\lab{pBFq}
\eeq
which corresponds to the $\alpha\beta$ part of \rf{BFYM0q}.
Notice that \rf{pBFq} is completely independent of the parameter $q$.

\subsubsection{The $BF$ theory with a cosmological term}
This theory is described by the action
\beq
S_{BF,\kappa} = i\braket B{*F_A} + i\frac\kappa2\braket B{*B},
\eeq
where $\kappa$ is a coupling constant known as the cosmological
constant. The symmetries are encoded in the following
BRST transformations:
\beq
\begin{array}{lclclcl}
sA &=& d_Ac +\kappa\tpsi, &\quad& sc &=& -\frac12[c,c]-\kappa\tphi,\\
sB &=& [B,c] - d_A\tpsi, & & & & \\
s\tpsi&=&-[\tpsi,c]+d_A\tphi, &\quad& 
s\tphi &=& [\tphi,c].
\end{array}
\lab{BRSTBFct}
\eeq
Again, $s^2$ is not nilpotent off shell and its failure
is given by
\[
s^2 B = -[F_A+\kappa B,\tphi]
\]
(notice that $[\tpsi,\tpsi]=0$).
The BV action reads
\beq
\begin{array}{lcl}
S_{BF,\kappa}^{\rm BV} 
&=& S_{BF,\kappa}+ 
\braket{d_Ac+\kappa\tpsi}{A^\dg}
+\braket{[B,c] - d_A\tpsi -\frac12*[B^\dg,\tphi]}{B^\dg} +\\
&+&\braket{-{1\over 2}[c,c]-\kappa\tphi}{c^\dg} +
\braket{-[\tpsi,c]+d_A\tphi}{\tpsi^\dg}+
\braket{[\tphi,c]}{\tphi^\dg} + 
\sum_i\braket{h^i}{\bc_i^\dg},
\end{array}
\lab{BFctBV}
\eeq
and the only field on which $\sigma$ acts in a different way is still 
$B$. $\sigma B$ is still given by \rf{sigmaBBF}, and its nilpotency
is ensured by
\[
\sigma B^\dg = [-B^\dg,c] - *(F_A +\kappa B).
\]

\paragraph{Quantization in the self-dual gauge}
The self-dual gauge is defined by putting $B^+=0$ and again
is well defined in the same hypotheses of Sec.\ \rf{sec-sd}.
Its quantum implementation is obtained by \rf{BVgf} with the
gauge-fixing fermion \rf{sdPsi} (forgetting of $\eta$).
The gauge-fixed action then turns out to be
\beq
S_{BF,\kappa}^{\rm s.d.\,g.f.} \begin{array}[t]{cl} 
= & -i\braket {B^-}{P^- F_A} - i\frac\kappa2 \braket {B^-}{B^-} +\\ 
+ & i\Big( s_0\PsiYM + 
\kappa\braket\tpsi{\frac{\delta\PsiYM}{\delta A}} + 
\braket\hone{D_A^*\tpsi} 
+ \hro i\braket{\omega_i[A]}\tpsi 
+  \braket{h_\chi^+}{B^+} + \\
+ & \braket\btphione{\Delta_A\tphi} + 
\bro i r^j \delta_{ij} \sqrt V 
+\kappa\braket\btphione{*[\tpsi,*\tpsi]} +
\kappa\,\bro i\braket\tpsi{\frac\delta{\delta A}\,
\braket{\omega_i[A]}\tpsi}+\\
+ & \frac12\braket{\bchi^+}{[\bchi^+,\tphi]}+
\braket{\bchi^+}{D_A\tpsi}
\Big),
\end{array}
\lab{SBFctgfsd}
\eeq
where $s_0$ is $s$ at $\kappa=0$.

Notice that $[\tpsi,\tpsi]$ vanishes, yet $[\tpsi,*\tpsi]$ does not.
This means that we have a source in $\btphione$; hence, the
$\btphione\tphi$-integration will produce a term which is quartic
in the ghost variables (vix., $\braket{[\bchi^+,\bchi^+]}
{*G_A*[\tpsi,*\tpsi]}$).

Therefore, differently from the covariant case \rf{BFYM-BFgf}, 
the BFYM theory and the $BF$ theory with a cosmological term
in the self-dual gauge have
quite different vertices. However, we can still relate their
quadratic parts.

\paragraph{Perturbative expansion}
We consider fluctuations as in \rf{fluct} with $q=\sqrt\kappa$,
$A_0$ an anti-self-dual-connection and $B_0=-F_{A_0}/\kappa$
(notice that this is a solution of the equations of motion in the
self-dual gauge). In this case, the quadratic action reads
\[
-i\braket{\beta^-}{d_{A_0}\alpha} + 
i\braket{F_{A_0}}{\alpha\wedge\alpha} -
\frac i2\braket{\beta^-}{\beta^-} + \mbox{gauge-fixing terms.}
\]
By making the change of variables
\[
\alpha' = e^{\frac{i\pi}4}\,\alpha,
\quad
\beta' = e^{-\frac{i\pi}4}\,\beta,
\]
we get
\beq
-i\braket{(\beta')^-}{d_{A_0}\alpha'} +
\braket{F_{A_0}}{\alpha'\wedge\alpha'} +
\frac12\braket{(\beta')^-}{(\beta')^-}
+ \mbox{gauge-fixing terms,}
\lab{BFtcq}
\eeq
which is exactly the $\alpha'\beta'$ part of \rf{BFYMsdq}.

\subsubsection{From BFYM to pure $BF$ theory as $\gym\to0$}
We want to find a canonical transformation that lets the symmetries
of BFYM theory become similar to those of $BF$ theory. First
we compute the action of $\sigma$ on $B^\dg$ in BFYM theory getting
\[
\sigma B^\dg = -[B^\dg,c]-*F_A+\sqrt2\,\gym\,d_A\eta
-2g^2B.
\]
Then we see that, if we make the change of variables
\beq
B\to \widetilde B = B + \sqtg *[B^\dg,\xi],
\lab{BBt}
\eeq
we get
\beq
\sigma\widetilde B = [\widetilde B,c] -d_A\tpsi
-*[B^\dg,\tphi] + *[d_A\eta,\xi]+[[B^\dg,\xi],\xi] 
-\sqrt2\,\gym *[\widetilde B,\xi].
\lab{sigmaBtilde}
\eeq
The transformation \rf{BBt} can be obtained as a canonical
transformation generated by
\beq
F(\Phi,\widetilde{\Phi^\dg}) =
\sum_i \braket{\Phi^i}{\widetilde{\Phi_i^\dg}} +
\frac1{2\sqrt2\,\gym}\,\braket{\widetilde{B^\dg}}
{*[\widetilde{B^\dg},\xi]}.
\eeq
Notice that, on all other fields than $B$ and on all antifields but
$\xi^\dg$, the transformation is the identity; on $\xi^\dg$
we have
\[
\xi^\dg\to\widetilde{\xi^\dg} = 
\xi^\dg - \frac1{2\sqrt2\,\gym}*\wcomm{B^\dg}{B^\dg}.
\]
Therefore, in the following we will drop all the tildes but on
$\widetilde B$ and $\widetilde{\xi^\dg}$.

We can now rewrite the BV action for BFYM theory in the new
variables:
\beq
\begin{array}{lcl}
\SBFYM^{\rm BV} &=& \widetilde\SBFYM + \braket{d_Ac}{A^\dg} +
\braket{[\eta,c] + d_A\xi - \sqrt2\, \gym\,\tpsi}{\eta^\dg} +\\
&+&\braket{[\widetilde B,c] - d_A\tpsi
-\frac12 *[B^\dg,\tphi] + *[d_A\eta,\xi]}{B^\dg}+\\
&+&\braket{
\frac12[[B^\dg,\xi],\xi] 
-\sqrt2\,\gym *[\widetilde B,\xi]}{B^\dg} +\\
&+&\braket{-{1\over 2}[c,c]}{c^\dg} +
\braket{-[\xi,c]+\sqrt2\, \gym\,\tphi}{\xi^\dg} +\\
&+&\braket{-[\tpsi,c]+d_A\tphi}{\tpsi^\dg}+
\braket{[\tphi,c]}{\tphi^\dg} + 
\sum_i\braket{h^i}{\bc_i^\dg},
\end{array}
\lab{newBVBFYM}
\eeq
where $\widetilde\SBFYM$ is the BFYM action evaluated at $\widetilde B$. 
Notice that now the BV action does not have singular 
terms in the limit $\gym^2\to0$; moreover, the BV
action is still proper in the limit.

Notice that, if we instead rescaled
$\xi\to\sqrt2\,\gym\,\xi,\widetilde{\xi^\dg}
\to\widetilde{\xi^\dg}/(\sqrt2\,\gym)$---in order for the BV transformation
\rf{sigmaBtilde} of $\widetilde B$ to become, in the limit
$\gym^2\to0$, the same as the BV transformation \rf{sigmaBBF} on $B$
in pure $BF$ theory---we would
not get a proper BV action in the limit.

The same problems would be encountered if we decided to rescale
also $\eta\to\sqrt2\,\gym\,\eta$ unless we introduced the required new
ghosts.

\paragraph{The partition function of BFYM at $\gym=0$}
Consider the $(\gym=0)$-BFYM action
\beq
\begin{array}{lcl}
\SBFYM^{\rm BV,0} &=& \widetilde\SBFYM^0 + \braket{d_Ac}{A^\dg} +
\braket{[\eta,c] + d_A\xi}{\eta^\dg} +\\
&+&\braket{[\widetilde B,c] - d_A\tpsi
-\frac12 *[B^\dg,\tphi] + *[d_A\eta,\xi]+
\frac12[[B^\dg,\xi],\xi]}{B^\dg} +\\
&+&\braket{-{1\over 2}[c,c]}{c^\dg} +
\braket{-[\xi,c]}{\xi^\dg} +\\
&+& \braket{-[\tpsi,c]+d_A\tphi}{\tpsi^\dg}+
\braket{[\tphi,c]}{\tphi^\dg} + 
\sum_i\braket{h^i}{\bc_i^\dg},
\end{array}
\eeq
Notice that the equations of motion impose $A$ to be flat.
Therefore, to quantize theory, it is convenient to
choose the covariant
gauge-fixing fermion $\Psi$ defined in \rf{Psi}.

After fixing the gauge,
we have at our disposal the rescaling
$\xi\to\epsilon\,\xi,\bxi\to\bxi/\epsilon$. Since the partition
function does not depend on the parameter $\epsilon$, we can
as well let $\epsilon\to0$. This way the $\eta,\bxi,h_\xi,\xi$
fields decouple from the others, and their contribution
to the partition function turns out to be
\[
\frac{\det\Delta_A^{(0)}}{({\det}'\Delta_A^{(1)}\,\det X_A)^{1/2}}.
\]
The $B$ integration then
selects the flat connections. The partition function of $BF$ theory
is the analytic torsion which is trivial in even dimension; 
moreover, notice that $X_A=1$ if $A$ is a flat connection.
Therefore, we have
\beq
\left.\ZBFYM\right|_{\gym^2=0} = \int_{A\in\calM_0} 
\frac{\det\Delta_A^{(0)}}{({\det}'\Delta_A^{(1)})^{1/2}},
\eeq
where $\calM_0$ is the moduli space of flat connections.

Notice that YM theory in the limit $\gym^2\to0$ leads to the same
result.

\section{Geometry}
\lab{sec-geo}
In this section we discuss the geometrical meaning of the set of fields
appearing in \rf{BRSTc-Aeta} and \rf{BRSTc-Btpsitphi}
and of the BRST equations \rf{BRSTBFYM}.\footnote{
To simplify the notations, we will take $\sqrt2\,\gym=1$ throughout
this section.}
The situation is as follows:
\begin{enumerate}
\item
In a topological gauge theory one deals with a connection $c$
on the bundle of gauge orbits $\calA\to \calA/\calG$, considers the 
corresponding
connection
$A+c$ on the $G$-bundle $P\times \calA$ and obtain  the BRST equations as  the
structure equations
and Bianchi identities for the curvature of $A+c$ \cite{BS}. 
\item In a non-topological Yang--Mills theory one 
considers the fiber immersion $j_A:\calG\to \calA$
and the pulled-back connection $j_A^*c$.
This is just the Maurer 
Cartan form on $\calG$;
the resulting structure equations for the curvature of
$A+j_A^*c$ give
the classical BRST equations\cite{BC}. It is customary to use the same symbol
$c$ also for the pulled-back connection $j_A^*c$: in Yang--Mills theory 
this is the ghost field.
\item In a full topological theory that includes the field 
$\eta\in\Omega^1(M,\ad P),$
one has to consider the tangent bundle $T\calA$ where there is a (free) action
of the tangent gauge group $T\calG$. In complete analogy to 
point 1.\ above, 
one should consider a connection on $T\calA$
and explicitly spell the structure equations and Bianchi identities
for the corresponding  connection on the $TG$-bundle $TP\times T\calA$. 
\item In a semi-topological theory that includes the field $\eta$,
one should {\em not} follow the analogy of point 2.\ above, i.e.,  
consider the $T\calG$-orbit in $T\calA$; instead one should
take into account 
the pulled-back bundle $j_A^* T\calA$ where $j_A:\calG\to \calA$
is the fiber immersion. The BRST equations will then be given as
the structure equations and the 
Bianchi identities for a curvature on the bundle
$TP\times j^*_AT\calA$.

In this way the connection $A$ is allowed to move only 
in a given $\calG$-orbit, as in the Yang--Mills theory, and
the only symmetry that the theory requires for the field $A$ is
gauge invariance. On the contrary, in the bundle $j^*_AT\calA$ the field 
$\eta$ may be any element of 
$\Omega^1(M,\ad P)\approx T_A\calA$; 
this means that the 
symmetries of the theory include the translation invariance for such an $\eta$.

In other words,
the theory is topological in one field-direction ($\eta$) and non-topological
in another field-direction ($A$).
\end{enumerate}
The last framework is the one that suits the theory described in this paper.
Another way to see this is to start with the action \rf{ABetagauge}
of the group $\Gaff$ on pairs $(A,\eta)\in T\calA$. Such an action does the
required job: it gauge-transforms $A$ and acts on $\eta$ by translations.
Unfortunately, such an action is not free,
so the quotient space is not a manifold. In order to turn around this 
problem, one has to consider exactly the bundle $j^*_AT\calA$ and 
obtain the BRST equations in the way mentioned above (point 4.) 
and discussed  in details in the following
pages. 

It is exactly in the framework of point 4.\ discussed  above
that  the field $B\in\Omega^2(M,\ad P)$ can
be included in the BRST equations by keeping the BRST operator nilpotent.

Dealing with the tangent gauge group and with the tangent bundle of the
space of connections means taking
first-order approximations. These relations can be made clearer if we consider
{\em paths (straight lines) of connections} on the bundle
$P^I\times \calA^I$, where $\calA^I\equiv Map([0,1],\calA)$ 
and $P^I$ is defined similarly.

Finally, in this section we are going to discuss the geometrical 
aspects of 
the gauge-fixing problems of our theory.

\subsection{Tangent gauge group}
\lab{subsec-tg}
Let $P$ be any $G$-principal bundle  over a closed oriented
manifold $M$.
The tangent bundle $TG$ of any Lie group is a Lie group itself 
which is
isomorphic to the semidirect product $G\times_s \lieg $ of the Lie group
with its Lie algebra. The product of two pairs $(g,x),(h,y)\in G\times_s
{\lieg}$ is defined as
\beq
(a,x)(b,y)\equiv (ab, \adb (x) +y).
\lab{productG}
\eeq
Its Lie algebra is the 
semi-direct sum of two copies of ${\lieg}$ with commutator
\beq
[(x_1,y_1),(x_2,y_2)]\equiv([x_1,x_2], [x_1,y_2]+[y_1,x_2]).
\lab{commutG}
\eeq
The tangent bundle $TP$ is a $TG$-principal bundle with base space  $TM$.
The action of $TG$ on $TP$ (obtained as the derivative of the $G$-action
on $P$) is given as follows
\beq
TP\times TG\ni (p,X)(g,x)\rsa \left(pg, (R_g)_*X, + i(x)_{pg}\right)\in
TP,
\lab{productTP}
\eeq
where $R_g$ denotes the (right) multiplication by $g\in G$ and $i(x)_p$
denotes the fundamental vector field corresponding to $x\in{\lieg}$
evaluated at $p\in P.$

A connection $A$ on $P$ is defined as 
a ${\lieg}$-valued one-form on $P$ with special
properties. First of all, we require its equivariance, viz.,
\[ 
A_{pg}\left((R_g)_*X\right) = \adg \left(A_p(X)\right), \quad X\in T_pP.
\]
Moreover, we require it to be the identity on fundamental vector fields:
\[
A_p(i(x)_p)=x,\quad \forall x\in {\lieg}.\]

A smooth map $\underline {p}: [a,b]^2\subset \reali^2\to P,$
with ${\underline p}(0,0)=p\in P$, defines an
element of the double tangent
\[
\left(p,{\underline p}^\prime,\dot{\underline p}, {d {\underline p}^\prime
\over dt}\right)\in TTP,\]
where $(t,s)\in [a,b]^2$ and the prime denotes the derivative w.r.t.\ $s$,
while the dot denotes the derivative w.r.t.\ $t$.

There is a canonical involution
\[
\alpha:TTP\to TTP,\quad \alpha\left(p,{\underline p}^\prime,\dot{\underline p},
{d {\underline p}^\prime
\over dt}\right)=\left(p,\dot{\underline p},{\underline p}^\prime, 
{d \dot{\underline p}
\over ds}\right).
\]
Now we consider the evaluation map
\[
ev:\calA \times TP \to {\lieg}
\]
and its derivative
\[ 
ev_* :T\calA \times TTP \to {\lieg} \times {\lieg}
\]
which has the following property
\begin{Thm}
For any $(A,\eta)\in T\calA$, the one-form on
 $TP$ given by
\beq
[\underline p]\rsa ev_*(A,\eta;\alpha_P[\underline p ])= \left(A_p\left(  
 \underline p' \right) ,\quad {d\over dt}\bigg|_{t=0} A_{\underline p(0,t) } 
( \underline p'(0,t) ) +
\eta\left( 
\underline{ \dot p} \right)\right)
\lab{AB}
\eeq
defines a connection on $TP$.
\lab{Thm-TPconn}
\end{Thm}
For the proof, s.\ \cite{CCR-tg}.
In this way we can identify the tangent bundle $T\calA$ as a subset of the 
space of connection on $TP$. It can be seen easily that it is a {\em proper}
subset \cite{CCR-tg}.

The gauge group $\calG$ is the space of equivariant 
maps
\[
\calG=Map_G(P,G)\ni g \Ra g(pa)=a^{-1}g(p)a ,\;\forall a\in G.\]
We have the following:
\begin{Thm}
The tangent gauge group $T\calG$ is a {\em proper subgroup} of the group
of gauge transformations for $TP$.
\end{Thm}

\paragraph{\em Proof}

Let $(\psi,\chi)\in\calG\times_s Lie(\calG)$.
Notice that for any $(p,X)\in TP$ and $(g,x)\in G\times_s \lieg$
we have the equation
\[
\begin{array}{ll}
\psi^{-1}d\psi\left[\left(pg, (R_g)_*X + i(x)|_{pg}\right)\right]=\\
{\Ad}_{g^{-1}}[\psi^{-1}d\psi (X)] +x -{\Ad}_{g^{-1}}
{\Ad}_{\psi^{-1}(p)} {\Ad}_{g}x.
\end{array}
\]
This shows that the map
\[(\psi,\chi):TP\lora G\times_s {\lieg}
\] given by
\[\left(\psi(p),
\psi^{-1}d\psi(p,X) + \chi(p) \right)\in G\times_s{\lieg}\]
is a gauge transformation for $TP$.
Notice that the above map is given by the derivative of the evaluation
map $ev:P\times \calG \to G$.

For any one-form $\eta\in\Omega^1(M,\ad P)$
and for any $(\psi,\chi)\in \calG\times_s Lie(\calG)$, the map
\[(p,X)\rsa \left(\psi(p), \psi^{-1}d\psi(p,X) + \chi(p) 
+ \eta(p,X) \right)\]
is also a gauge transformation on $TP$, thus showing that the inclusion
$T\calG_P\subset \calG_{TP}$ is proper.
\QED

In the proof of the previous theorem we showed explicitly that the group 
$\Gaff$
(the semidirect product of $T\calG$ with the abelian group $\Omega^1(M,\ad P))$
is also a subgroup of $\calG_{TP}$.
\begin{Rem}
{}From the discussion above we conclude that $T\calG$ acts freely on $T\calA$
and this action coincides with the restriction of the action of the gauge group
of  $TP$ ($\calG_{TP}$) on the space
$\calA_{TP}$ of connections on $TP$.
\end{Rem}
\begin{Rem} The group $\Gaff$ acts non-freely on $T\calA$ as in 
\rf{etagauge}. The group $\Gaff$ is  a subgroup
of $\calG_{TP}$, but the action \rf{etagauge} is {\em not} given by the
restriction of the action of $\calG_{TP}$ on $\calA_{TP}$.
\end{Rem}

\subsection{Paths on a principal bundle}\lab{subsec-paths}
For any manifold $X$ we denote by $X^I$ the space of smooths paths
$Map(I,X)$ where $I=[0,1]$.
If $P(M,G)$ is a principal bundle,
the group $G^I$ acts freely on $P^I$ and the bundle $P^I(M^I,G^I)$ is a
principal bundle.

A path in $\calA\equiv\calA_P$ defines 
a connection on $P^I$. In this way we identify
$\calA^I$ with $\calA_{P^I}$. 

There is a natural bundle homomorphism
\beq
P^I\to TP,\quad p(t)\rsa \left(p(0), \dot p(0)\right)
\lab{pathP}
\eeq
which corresponds to the group homomorphism
\beq
G^I\to G\times_s\lieg, \quad g(t) \rsa \left(g(0), g^{-1}(0)\dot g(0)\right).
\lab{pathG}
\eeq
Under the homomorphisms \rf{pathP} and \rf{pathG}, a connection $A(t)$
is sent into $\left(A(0),\dot A(0)\right)\in T\calA.$

If we have a connection $c$ (a.k.a.\ as a gauge fixing) on the bundle of
gauge orbits
$\calA\to \calA/\calG$, then $A+c$ is a connection on
 the bundle
\beq
{{P\times \calA}\over{\calG}}\mapsto M\times {\calA\over \calG}.
\lab{peg}
\eeq
In fact 
the one-form on $P\times \calA$ given by
\beq
(A+c)_{(p,A)}(X, \eta)\equiv A(X)_p + c(\eta)_{(A,p)}\lab{connpeg}
\eeq
is a connection on $P\times \calA$ which is $\calG$-invariant, i.e.,
descends to a connection on the principal $G$-bundle \rf{peg}.

Forms on $P\times \calA$ have a bi-degree $(k,s)$ where $k$ is the order 
of the form on $P$ and $s$ is the order of the form on $\calA$, 
a.k.a.\ the {\em ghost number}. 

By taking the tangent bundles of  \rf{peg} one obtain the bundle
\beq
{{TP\times T\calA}\over{T\calG}}\mapsto TM\times {T\calA\over T\calG}.
\lab{tpetg}
\eeq
By considering the relevant path spaces, one has the bundle
\beq
{{P^I\times \calA^I}\over{\calG^I}}\mapsto M^I\times {\calA^I\over \calG^I}.
\lab{pathpeg}
\eeq
If $c(t)$ is a path of connections in $\calA\to\calA/\calG$, and $A(t)$ is 
a path of connections in $\calA$, then a connection on \rf{pathpeg}
is given by 
\beq
A(t)+c_{A(t)}(t),
\lab{connpath}
\eeq
where we have explicitly represented the dependency of the connection $c(t)$
on the point $A(t)\in\calA$.

As particular paths we can take straight lines,
\beq
A(t)=A+t\eta,\quad\eta\in\Omega^1(M,\ad P),\quad\quad c(t)=c+t\hat c,
\lab{aetacc}
\eeq
where $\hat c$ is an assignment to each connection $A\in \calA$ of a map
${\hat c}_A:\Omega^1(M,\ad P)\mapsto \Omega^0(M,\ad P)$ with the property
of $\calG$-equivariance,
\[{\hat c}_{A^g}\left({\Ad}_{g^{-1}}\tau\right)
= {\Ad}_{g^{-1}}\left({\hat c}_A(\tau)\right),\]
and of tensoriality,
\[\hbox{Im}\,(d_A)\subset \ker ({\hat c}_A).\]
In physics, ${\hat c}$ is an {\it infinitesimal variation of the 
gauge fixing}.
It is convenient to rewrite the connection given by \rf{aetacc} as
\beq 
A+t\eta + c_{A+t\eta} +t \hat c_{A+t\eta} = A + c_{A} + t 
\left(\eta +\xi_{A,\eta}\right) + \sum_{n=2}^{+\infty}t^n\xi_{A,\eta}^{(n)}.
\lab{linearconn}
\eeq
In the previous expression we have:
\begin{enumerate}
\item identified the tangent bundle 
$T\calA$ with $\calA\times\Omega^1(M,\ad P)$ [forms on $\calA$
can then be evaluated  on elements of $\Omega^1(M,\ad P)$];
\item 
defined \[\xi^{(n)}_{A,\eta}(\tau)\equiv \displaystyle{{1\over n!}
{d^n\over dt^n}\bigg|_{t=0} c_{A+t\eta}(\tau)
+{1\over{(n-1)!}}{d^{(n-1)}\over dt^{n-1}}\bigg|_{t=0}\hat c_{A+t\eta}(\tau)},
\quad \tau\in\Omega^1(M,\ad P),\]
and set $\xi_{A,\eta}\equiv\xi^{(1)}_{A,\eta}.$
\end{enumerate}
As will be shown in a moment,
the pair $(A+c,\eta + \xi)$ can be seen as an honest connection with values in
the Lie algebra of 
the tangent group $TG$.

First we notice that an infinite-dimensional version 
of \rf{Thm-TPconn} implies that the pair $(c,\hat c)$  defines a connection
on the $T\calG$-bundle $T\calA$.
Explicitly we have:
\begin{Thm}
When we identify the double tangent bundle $TT\calA$ with $\calA\times
\Omega^1(M,\ad P)^{\times 3}$, then the connection on $T\calA$
represented by $(c,{\hat c})$ is a map
\[\calA\times
\Omega^1(M,\ad P)^{\times 3}\ni(A, \eta, \tau, \sigma)\rsa \left(c_A(\tau), 
\xi_{A,\eta}(\tau)
+c_A(\sigma)\right)\in \lieg \oplus_s \lieg.\]
\lab{th-chatc}
\end{Thm}

Now we look again at the bundles
\rf{tpetg} and \rf{pathpeg}. 

Given the  natural inclusions
\[P\to TP, p\rsa(p,0), \quad P\to P^I, p\rsa [p(t)=p],\]
we establish from now on the following
\begin{Con}
We will generally 
assume that the forms on $TP\times T\calA$ we are going to consider
are restricted 
to forms on $P\times T\calA$ and that the forms 
on $P^I\times \calA^I$ we are going to consider 
are restricted to forms on $P\times \calA^I$.
\end{Con}

Moreover, we assume
\begin{Con} 
We consider only elements in $TT\calA\approx 
\calA\times\Omega^1(M,\ad P)^{\times 3}$ that have $0$ as fourth component.
\end{Con}

We conclude that the $\lieg$-valued form 
\beq
(A+c,\eta + \xi),
\lab{acex}
\eeq
whose explicit expression is given by
\[(A+c,\eta+\xi)_{p;A,\eta}(X,\tau)=
\left(A_p(X)+c_A(\tau),\eta_p(X)+\xi_{A,\eta}(\tau) \right),\]
with $p\in P, X\in T_pP $ and $(A,\eta,\tau)\in \calA\times
\Omega^1(M,\ad P)^{\times 2}$,
represents a 
connection on the bundle \rf{tpetg} provided that {\em conventions}
1 and 2 are understood. 

\subsection{Curvatures}
\lab{subsec-curv}

As is customary in topological (cohomological) field theories
\cite{BS},
the BRST equations are nothing but the structure equations and the Bianchi
identities for the connections of some bundles of fields.

We start by recalling the expression
of  the curvature of the connection \rf{connpeg}. It is given by
\beq
F_A + \psi +\phi,
\lab{curvpeg}
\eeq
where the three terms above are forms of degree $(2,0)$, $(1,1)$, $(0,2)$
in the product space $P\times \calA$, the second number being the 
ghost number.

More precisely:
\begin{enumerate}
\item $\psi$ is minus the projection of $\Omega^1(M,\ad P)$ on the horizontal
subspace;
\item $\phi$ is the curvature of the connection $c$ on the bundle $\calA
\to \calA/\calG$.
\end{enumerate}.

The structure equations and the Bianchi identities in this case read
\beq
\begin{array}{lclclclclcl}
\delta A&=& d_A c - \psi,&\quad& \delta c &=& -{1\over 2}[c,c] +
\phi, &\quad& d_AF_A&=&0,\\
\delta \psi&=& d_A\phi -[\psi,c], &\quad& \delta\phi&=&[\phi,c]
&\quad& \delta F_A &=&[F,c]-d_A\psi,
\end{array}
\lab{topgauge}
\eeq
where we have denoted by
$\delta$ the exterior derivative on $\calA$, a.k.a.\ as the BRST
operator. The total derivative for $(k,s)$-forms on $P\times \calA$
is given by
\beq
d_{tot} = d + (-1)^k \delta.
\lab{dtot}
\eeq
The commutator of forms of bidegree $(k,s)$
is assumed to satisfy the equation \[
[\omega^{(k_1,s_1)}_1,\omega^{(k_2,s_2)}_2]=(-1)^{k_1k_2 + s_1s_2
+1}[\omega^{(k_2,s_2)}_2,\omega^{(k_1,s_1)}_1].
\]
In this way the total covariant derivative satisfies the same sign-rule
as \rf{dtot}, namely it is given by $d_A + (-1)^k \delta_c.$

Equations \rf{topgauge} are the field equations for the topological field 
theory
considered in \cite{BS}.

Next we consider the curvature of \rf{linearconn}. It is given by
\beq
F_A + \psi + \phi + t\left (d_A\eta + \tilde\psi + \tilde\phi\right) + t^2 
\left(\tilde\psi^{(2)} + \tilde\phi^{(2)} +{1\over 2}
[\eta + \xi,\eta + \xi]\right) + o(t^2),
\lab{linearcurv}
\eeq
where the forms $\tilde\psi$ and $\tilde\phi$ are defined as
\beq
\begin{array}{lcl}
\tilde \psi&\equiv& d_A\xi +[\eta,c]-\delta \eta ,\\
\tilde \psi^{(2)}&\equiv& d_A\xi^{(2)},\\
\tilde \phi&\equiv&\delta \xi + [c,\xi].\\
\tilde \phi^{(2)}&\equiv&\delta \xi^{(2)} + [c,\xi^{(2)}]=\delta_c\xi^{(2)}
\end{array}
\lab{tildes}
\eeq
Accordingly the curvature of the connection \rf{acex} is given by the 
first-order 
term of \rf{linearcurv}, i.e., by the $\lieg\oplus_s\lieg$-valued
form
\beq
(F+\psi +\phi, d_A\eta +\tilde\psi+\tilde\phi).
\lab{curvacex}
\eeq
The structure equations and the Bianchi identity for \rf{curvacex} and 
\rf{linearcurv} are the natural generalization of \rf{topgauge}. They are
spelled out explicitly  in \cite{CCR-tg}. 
Here we are concerned with the geometrical interpretation of the BRST 
equations considered in Sec.\ \ref{sec-BFYM},
and this requires a restriction to the
$\calG$-fiber in the bundle $T\calA\to T\calA/T\calG$.

\subsection{Restriction to the $\calG$-fiber}
\lab{subsec-restr}
In order to obtain the standard BRST equation from \rf{topgauge}, 
we need 
to consider the fiber imbedding
\beq
j_A:\calG\to \calA,\quad j_A(g)=A^g,
\lab{ja}
\eeq
and the pulled-back bundle 
\beq
P\times \calG\to M\times \calG.
\lab{YMPG}
\eeq
The connection $A+c$ \rf{connpeg} on $P\times \calA$ is pulled back to
\rf{YMPG}. It is customary (and unfortunately confusing)
to denote the pulled back connection by the same symbol $A+c$. This means
that in this case {\em  $c$
is just the Maurer--Cartan form on} $\calG$.
The curvature of the pulled-back connection is simply $F_A$, and the structure
and Bianchi identities become
\beq
\delta A= d_A c, \quad \delta c = -{1\over 2}[c,c], \quad d_AF_A=0,\quad 
\delta F_A =[F_A,c].
\lab{YMgauge}
\eeq
These are the standard BRST equation for the Yang--Mills theory
and the  connection $A+c$ on \rf{YMPG}
gives the geometrical interpretation of the set 
of fields and ghosts appearing in \rf{BRSTc-YM}
\cite{BC}.

Let us apply now a similar fiber imbedding to the bundle
$TP\times T\calA$. Two choices are possible:
\begin{enumerate}
\item consider the fiber imbedding of the full tangent gauge group $T\calG$, 
i.e., the bundle:
$TP\times j_{A,\eta}T\calG \to TM \times j_{A,\eta}T\calG$
\item consider only the fiber imbedding \rf{ja} and restrict the
tangent bundle $T\calA$ to the image of $j_A$. This means
 considering the pulled-back bundle
\beq
TP\times j_A^*(T\calA)\to TM \times j_A^*(T\calA).
\lab{jstar}
\eeq
\end{enumerate}
The first alternative 
would lead us to dealing with the Maurer--Cartan form on
$T\calG$. But we are in fact interested in the second alternative 
since we
want the field $\eta$ to be generic and not
restricted to be tangent to the $\calG$-orbit. 
Hence, from now on, only the {\em second} alternative will be 
considered:
this means that in the connection
$(A+c,\eta +\xi)$ the form $c$ becomes the Maurer--Cartan form
on $j_A(\calG)$.

Taking always into consideration
{\em convention} 1, the corresponding curvature becomes
\beq
(F_A, d_A\eta +\tilde\psi+\tilde\phi).
\lab{fibercurvacex}
\eeq
The Bianchi and structure equation for \rf{fibercurvacex} are
\beq
\begin{array}{lcl}
\delta A&=& d_A c,\\
\delta \eta&=&-\tilde \psi + d_A\xi +[\eta,c],\\
\delta F_A&=&[F_A,c],\\
\delta(d_A\eta)&=& -d_A\tilde\psi + [F_A,\xi]
+[d_A\eta,c],\\
\delta c &=& -{1\over 2} [c,c],\\
\delta \xi&=&\tilde \phi - [c,\xi],\\
\delta \tilde \psi&=& -[\tilde \psi, c] + d_A\tilde\phi,\\
\delta \tilde \phi &=&[\tilde \phi, c].
\end{array}
\lab{fiberequations}
\eeq
{\em The connection $(A+c,\eta+\xi)$ for  
\rf{jstar}  gives the geometrical interpretation of the set 
of fields and ghosts appearing in \rf{BRSTc-Aeta}
and \rf{BRSTc-Btpsitphi}} (but for the field $B$).

The analogous construction for the connection
\rf{linearconn} implies the following steps:
\begin{enumerate}
\item take the map 
\beq
\begin{array}{ccc}
j_A^*(T\calA)\approx \calG\times T_A\calA & \to &\calA^I,\\
(g,A,\eta)& \rsa &(A+t\eta)^g;
\end{array}
\lab{fiberlinear}
\eeq
\item pull back the connection \rf{linearconn} to the bundle
\[
P^I\times \calG\times T_A\calA; 
\]
\item consider only {\em constant} paths in $P^I$, according to 
{\em convention} 1.
\end{enumerate}
At this point the formal expression of the connection is the same
as in \rf{linearconn}, i.e.,
\beq
A + c + t 
\left(\eta +\xi\right) + \sum_{n=2}^{+\infty}t^n\xi^{(n)},
\lab{fiberlinearconn}
\eeq
but now $c$ is the Maurer--Cartan form.

The relevant curvature is
\beq
F_A  + t\left (d_A\eta + \tilde\psi + \tilde\phi\right) + t^2 
\left(\tilde\psi^{(2)} + \tilde\phi^{(2)} +{1\over 2}
[\eta + \xi,\eta + \xi]\right) + o(t^2).
\lab{fiberlinearcurv}
\eeq

Computing the structure and Bianchi identities for \rf{fiberlinearcurv}
will give again \rf{fiberequations} and some other transformation
laws for the fields 
$\tilde\psi^{(n)},\tilde\phi^{(n)},\tilde \xi^{(n)}$
which we do not discuss here (s.\ \cite{CCR-tg}). 

\subsection{Including the field $B$}
\lab{subsec-Bfield}

In four-dimensional 
$BF$ quantum field theories, the field $B$
behaves like a curvature but does not depend on the connection.
It is then represented
by an element
of $\Omega^2(M,\ad P)$.

Now we show that such a field can be incorporated into the field
equations \rf{fiberequations}.
By incorporating we mean that the BRST double complex with operators
$(d,\delta)$
can consistently be
extended to a double complex with operators $(d,s)$
that includes the space $\Omega^2(M,\ad P)$ in a such a way that:
\begin{enumerate}
\item $s$ extends $\delta$, so that $s^2=0$; 
\item the gauge equivariance is preserved, and 
\item the field equations are preserved.
\end{enumerate}

We use here the same notation of section 2.; viz., we set
\[
\calB\equiv \Omega^2(M,\ad P),
\]
and consider the  tangent bundle
\[T\calB\approx \Omega^2(M, \ad P)\times \Omega^2(M,\ad P).
\]
The group $T\calG$ acts on the cartesian product
$T\calA \times T\calB$ as follows:
\beq
\begin{array}{cl}
(A,\eta;C, E)\cdot (g,\zeta)=\\
\left( A^g, {\Ad}_{g^{-1}}\eta
+d_{A^g} \zeta; {\Ad}_{g^{-1}} C, {\Ad}_{g^{-1}}E +
[{\Ad}_{g^{-1}} C,\zeta]
\right),
\end{array}
\lab{tgontatb}
\eeq
yielding a principal $T\calG$-bundle.

Since the projection $\calA\times \calB\mapsto \calA$ is a morphism
of $\calG$-bundles, the connection $c$ on $\calA$
is also a connection
on $\calA\times\calB$, and the  connection $(c,\xi)$ [determined by the
pair $(c,\hat c)$] on $T\calA$
is also a connection on $T\calA\times T\calB$.

Moreover, $(A+c, \eta+ \xi)$ is a 
$Lie(TG)$-valued {\it connection} on the bundle
\beq
{{TP\times T\calA\times T\calB}\over{T\calG}}\mapsto TM\times {T\calA\times
T\calB \over T\calG},
\lab{tptatb}
\eeq
where again we intend to apply  {\em conventions} 1 and 2.

Forms on $TP\times T\calA\times T\calB$ will be characterized by
three indices $(m,n,p)$ which represent the degree with respect to the
three spaces $TP$, $T\calA$, $T\calB$. The middle integer $n$
is again the ghost number.

The pair $(C,E)\in T\calB$ is a $Lie(TG)$-valued $(2,0,0)$-form that
is constant on $T\calA.$

If we 
neglect the forms of degree $(m,n,p)$ with $p>0$,
we find that,
under the action of
the total covariant derivative $d^{tot}_{(A+c,\eta+\xi)}$, the pair 
$(C,E)$ is transformed into
\[
\left(d_AC + [c,C],d_A E + [\eta,C] + [c,E] +
[\xi,C]
\right),
\]
where we have used the fact that the pair $(C,E)$ is constant
on the space $T\calA$.

Now we are ready to consider the possible extensions of the BRST 
operator
$\delta$ to $T\calB$ that satisfy the requirements 1, 2 and 3 above.
In order to take into account requirement number 2, we have to 
consider covariant derivatives, so 
we set
\beq
s(C,E)\equiv ([C,c], [C,\xi] +[E,c]);
\lab{esse}
\eeq
i.e., $(d_A - s)(C,E)$ coincides, up to forms of positive degree in the 
$T\calB$-component, with
$d^{tot}_{(A+c,\eta+\xi)}(C,E)$.

If the $(2,2,0)$ component
of \beq
\left(d^{tot}_{(A+c,\eta+\xi)}\right)^2(C,E)=[F_{(A+c,\eta+\xi)},(C,E)]=
([F_A,C],[F_A,E]+[d_A\eta+\tpsi+\tphi,C])
\lab{commutF}
\eeq
is zero, then we may add to
our field-equations \rf{fiberequations} the transformations \rf{esse}
and obtain a consistent
BRST algebra that includes the elements $(C,E)\in T\calB$.

It is a matter of simple calculations to check
the following
\begin{Thm} A consistent BRST algebra that includes pairs
$(C,E)\in T\calB$ and extends \rf{fiberequations} is possible only
for pairs $(0,E)$ for any $E$.
\end{Thm}

If we perform the change of variables
\beq
d_A\eta + E = B
\lab{change}
\eeq
and replace $\delta$ with $s$ in \rf{fiberequations},
we obtain the following set of equations:
\beq
\begin{array}{lcl}
s A&=& d_A c,\\
s\eta&=&-\tilde \psi + d_A\xi +[\eta,c],\\
s F_A&=&[F_A,c],\\
s B&=& -d_A\tilde\psi + [F_A,\xi]
+[B,c],\\
s c &=& -{1\over 2} [c,c],\\
s \xi&=&\tilde \phi - [c,\xi],\\
s \tilde \psi&=& -[\tilde \psi, c] + d_A\tilde\phi,\\
s \tilde \phi &=&[\tilde \phi, c],
\end{array}
\lab{Bfiberequations}
\eeq
which are immediately recognized as the equations \rf{BRSTBFYM}.

The change of variables \rf{change} implies that $B$ is a tangent vector 
in $T_{F_A}\calB$. Accordingly its transformation under the group $T\calG$
is as follows:
\beq
B\cdot(g,\zeta)=\Ad_{g^{-1}}B + [F_{A^g},\zeta].
\lab{tgB}
\eeq

\subsection{Gauge fixing and orbits}
\lab{subsec-gf}
The fields of our theory are triples
$(A,\eta,B)$, where $(A,\eta)\in T\calA$ and $B\in T_{F_A}\calB$. 
This space of fields can be described conveniently by means of
the curvature map
\beq
K:\calA\to\calA\times\calB,\quad K(A)= (A,F_A),
\lab{kappa}
\eeq
which descends to a map
\[
K:{\calA\over\calG}\to {{\calA\times\calB}\over \calG}.\]
The space of fields [i.e., the set of triples $(A,\eta,B)$]
coincides then with the set of elements of the
pulled-back bundle
\[ K^*(T\calA\times T\calB).
\]
By taking into account the $T\calG$-invariance, the space of orbits of the 
theory is given by
\beq
{K^*(T\calA\times T\calB)\over T\calG}\approx {K^*(H\calA\times
T\calB)\over\calG},
\lab{KHA}
\eeq
where by $H\calA$ we denote the bundle of
horizontal tangent vectors of $T\calA$
with respect to a given connection on 
$\calA\to (\calA/\calG)$. 
The above diffeomorphism is induced by
the linear map
\[
[A,\eta,B]_{T\calG}\,\rsa\,
[A,\eta^H,B-d_A\eta^V]_\calG,
\]
where the superscript $H$ and $V$ denote the horizontal and vertical 
component.

The general $\Gaff$-invariance of the
action \rf{SBFYM} implies the following further
translational invariance:
\beq
K^*(H\calA\times T\calB)\ni (A,\eta,B)\rsa (A,\eta+\tau, B-d_A\tau),
\quad \tau\in H\calA.
\lab{transl-invar}
\eeq

Relatively to this translational invariance, 
two different gauges are possible:
\begin{enumerate}
\item $\eta=0$;
\item $B\perp d_A(H\calA)$; i.e., $d_A^*B\in {\rm Im}(d_A)$.
\end{enumerate}

We can therefore identify the space of gauge-fixed fields as
\beq
{K_2^* T\calB\over\calG}\approx{ {\left(H\calA\ominus {\rm Harm}_A^1(M,\ad P)
\right)
\oplus\hat \calB}\over\calG},
\lab{gaugefixedfields}
\eeq
where $K_2:\calA\to\calB$ denotes the second component of $K$, and $\hat\calB$
is a vector bundle over $\calA$ defined by:
\[
\hat
\calB_A\equiv\{B\in \Omega^2(M,\ad P)\ |\ d_A^*B\in
{\rm Im}(d_A)\}.
\]

Let us finally come to the self-dual gauge. Here we consider
the operator 
$D_A:H\calA\to \Omplus$ 
[s.\ \rf{deTA}], and assume the following condition:
\beq
{\rm Im}(D_A)=\Omplus.
\lab{surje}
\eeq

We know that \rf{surje} is satisfied when 
$A$ is an anti-self-dual connection.
By the same argument discussed in subsec.\ \rf{sec-sd}, we conclude that, 
if $A$ is a connection such that \rf{surje}
is satisfied, then 
there is a neighborhood of $A$ in which the same condition is satisfied 
as well;
so the set of connections for which \rf{surje} is satisfied is an open set.  

On such an open set there is another way of fixing the 
translational invariance \rf{transl-invar}.
If we set 
\beq B^+=0 \Lra B\in \Omminus  \Lra B\perp \Omplus,
\lab{Bplus}
\eeq
then $\tau\in H\calA$ is determined only 
up to elements in $\tHarm_A^1(M,\ad P)$.
We can then conclude that  the space
of gauge-fixed fields can, in a neighborhood of  
an anti-self-dual connection, be given by
\beq
{K_2^* T\calB\over\calG}\approx{ {\left(H\calA\ominus \tHarm_A^1(M,\ad P)
\right)
\oplus\Omminus}\over\calG}.
\lab{gaugefixedfieldsbis}
\eeq

\section{Conclusions}
In this paper we have discussed the possibility of describing YM
theory in terms of a theory that shares many characteristics with
the topological field theories (of the $BF$ type).

The first step, Sec.\ \ref{sec-prem}, 
has been considering the first-order formulation
of YM theory with the addition of an extra field to be gauged away.
The resulting theory \rf{SBFYM}, which we call BFYM theory, shows
a formal resemblance with pure $BF$ theory as the coupling constant
vanishes.

In Sec.\ \ref{sec-BFYM}, we have shown that BFYM theory is indeed
equivalent to YM theory. Our proof is an explicit path-integral
computation performed with three different (but equivalent) gauge 
fixings: the trivial, the covariant and the self-dual.
The most interesting result is that perturbation theory in the
last two gauges can be organized
in a different way than in the second-order YM theory and explicitly shows
the propagators of the topological $BF$ theories.

In Sec.\ \ref{sec-BV}, after recalling some basic facts on the BV
formalism, we have given a brief description of the BV quantization
of the $BF$ theories. Moreover, we have shown that BFYM theory
can be formulated in a canonically equivalent way so that the limit
for vanishing coupling is well-defined and yields the pure $BF$ theory
plus a covariant kinetic term for the extra field.

Finally, in Sec.\ \ref{sec-geo}, we have described the geometric
structure of BFYM theory and have explicitly shown how to deal with
the non-freedom of the action of the symmetry group on the space of
fields.

We conclude by recalling that one of the reasons for considering
BFYM theory (and looking for its underlying topological properties)
is the possibility of introducing new observables that might
realize `t~Hooft's picture; but this will be discussed elsewhere.

\section*{Acknowledgments}
Two of us (F.F. \& A.T.) thank S.P. Sorella for
useful discussions.

\appendix

\section{Computation of the Pfaffian of $M$}
\lab{app-Pf}
When studying BFYM theory in the self-dual gauge, we needed
to compute the Pfaffian of the matrix $M$ defined in \rf{defM}.
By a well known algebraic identity,
\[
({\rm Pf}M)^2 = \det M;
\]
therefore,
\[
({\rm Pf}M)^4 = \det M^2.
\]
An explicit computation using \rf{tDeltaA'} and
the last line of \rf{tDelta} yields
\[
M^2 = -\left(\begin{array}{ccc}
\tDelta_A' & 0 & 0\\
0 & \sqrt V {\bf 1} & 0\\
0 & 0 & N
\end{array}\right),
\]
with
\[
N = \left(\begin{array}{cc}
\frac12\tDelta_A & D_AD_A\\
D_A^*D_A^* & \tDelta_A
\end{array}\right)
\]
acting on $\Omplus\otimes\Omega^0(M,\ad P)$. Up to a possible
irrelevant phase, the determinant of $M^2$ is equal to the product
$V^{m^-/2}\,{\det}'\tDelta_A^{(1)}\,\det N$.  An explicit computation
yields
\[
\det N = \det(\tDelta_A^{(2)}/2)\, \det(\tDelta_A^{(0)} - R_A),
\]
with
\[
R_A = D_A^*D_A^*2\tG_AD_AD_A : \Omega^0(M,\ad P)\to \Omega^0(M,\ad P).
\]
To simplify $R_A$, we notice that
\[
2\tG_AD_A : \Omega^1(M,\ad P)\to\Omplus
\]
is the inverse of $D_A^*$ on its image; more precisely, we have
\[
D_A^*2\tG_AD_A = \pi_{{\rm coker}(D_A)},
\]
which implies \rf{defR}.

Moreover, by the last line of \rf{tDelta}, we have
\[
\det(\tDelta_A^{(2)}/2) = \det(\Delta_A^{(2,+)}).
\]

Putting together all the pieces, we finally get \rf{Pf}.

\section{Some useful properties of the operator $D_A$}
\lab{sec-pq}
In Sec.\ \ref{sec-sd}, we have introduced the operator $D_A$.
This operator is an injection from the zero- to the one-forms
since $A$ is irreducible.
Moreover, we have supposed to work in a neighborhood $\calN'$ of the
space of anti-self-dual connections where $D_A$ is also a 
surjection from the one- to the self-dual two-forms.

These two properties are enough to prove a series of
facts which were used in Sec.\ \ref{sec-sd} to prove the
equivalence between YM theory and BFYM theory in the self-dual gauge.

In subsection \ref{sec-genc} we will prove these facts in general;
in subsection \ref{sec-ourc} we will specialize to our case.

\subsection{The general case}
\lab{sec-genc}
Let us consider three (finite-dimensional)
vector spaces $X$, $Y$ and $Z$ together
with an injective linear operator
\beq
p : X\to Y
\eeq
and a surjective linear operator
\beq
q : Y\to Z.
\eeq
Moreover, we assume ${\rm ker}\,q\bigcap{\rm ker}\,p^*=\{0\}$.

Then we introduce the Laplace operators
\beq
\begin{array}{lcl}
\Delta_X &=& p^*p,\\
\Delta_Y &=& pp^* + q^*q,\\
\Delta_Z &=& qq^*.
\end{array}
\eeq
With our hypotheses, these operators are invertible; we will denote
by $G_*$ their inverse.

Now we have the following:
\begin{Thm}
If $qp=0$, then
\begin{enumerate}
\item $0\to X \stackrel p\to Y \stackrel q\to Z \to 0$ is
an exact sequence, and
\item 
\[
\frac{\det\Delta_X\,\det\Delta_Z}{\det\Delta_Y} = 1.
\]
\end{enumerate}
\lab{thm-qp0}
\end{Thm}

\paragraph{\em Proof} 
Fact 1 just follows from the definition of exact sequence. For fact
2, write
\[
Y = Y_1 \oplus Y_2,
\]
with
\[
Y_1 = {\rm Im}\,p={\rm ker}\,q, \quad Y_2 = {\rm Im}\,q^*={\rm coker}\,q.
\]
$\Delta_Y$ is block diagonal with respect to this
decomposition, so $\det\Delta_Y = \det\Delta_{Y_1}\,\det\Delta_{Y_2}$,
with
\[
\begin{array}{ccccc}
\Delta_{Y_1} &=& pp^* &:& Y_1\to Y_1,\\
\Delta_{Y_2} &=& q^*q &:& Y_2\to Y_2.
\end{array}
\]
Moreover, $Y_1$ ($Y_2$) is isomorphic to $X$ ($Z$), and
$p$ and $p^*$ ($q^*$ and $q$) are invertible operators when restricted
to this space. It follows that
\[
\det\Delta_{Y_1} = \det\Delta_X, \quad
\det\Delta_{Y_2} = \det\Delta_Z.
\]
\QED

If $pq\not=0$, we cannot identify ${\rm ker}\,q$ with ${\rm Im}\,p$.
We can
however reduce to the preceding situation by defining
\beq
\bar p = Hp : X\to Y,
\eeq
where $H$ is the projection operator
\beq
H = 1-q^*G_Zq : Y\to Y.
\eeq
Since $qH=0$, we get $q\bar p=0$. We will then define
\beq
\begin{array}{lcl}
\bar\Delta_X &=& \bar p^*\bar p,\\
\bar\Delta_Y &=& \bar p\bar p^* + q^*q.
\end{array}
\eeq
Then we have the following
\begin{Cor}
If $\bar p$ is injective and $q$ is surjective, then
\[
\frac{\det\bar\Delta_X\,\det\Delta_Z}{\det\bar\Delta_Y} = 1.
\]
\end{Cor}

This corollary can however be refined to give the following
\begin{Thm}
If $\bar\Delta_X$ is invertible and $q$ is surjective, then
\[
\frac{\det\bar\Delta_X\,\det\Delta_Z}{\det\Delta_Y} = 1.
\]
\lab{thm-bar}
\end{Thm}

\paragraph{\em Proof}
First of all we notice that ``$\bar p$ injective" is equivalent to 
``$\bar\Delta_X$ invertible."  

Then we proceed as in the proof of Thm.\ \ref{thm-qp0} and split
$Y$ as
\[
Y = Y_1 \oplus Y_2,
\]
where $\bar Y_1 = {\rm ker}\,q$ and $Y_2={\rm coker}\,q$. Notice
that $HY_2=\{0\}$ and $H|_{Y_1}=1$, so ${\rm ker}\,H=Y_2$.

The Laplace operator $\bar\Delta_Y$ is block diagonal with respect
to the above decomposition of $Y$. Therefore, 
\[
\det\bar\Delta_Y = {\det}_{Y_1}(pp^*)\,{\det}_{Y_2}(q^*q).
\]

The Laplace operator $\Delta_Y$ is not block diagonal but has
the following form with respect to the above decomposition:
\[
\Delta_Y = \left(\begin{array}{cc}
pp^* & pp^* \\
pp^* & pp^*+q^*q
\end{array}\right).
\]
By subtracting the first from the second row, we get
\[
\det\Delta_Y = \det\left(\begin{array}{cc}
pp^* & pp^* \\
0 & q^*q
\end{array}\right) = \det\bar\Delta_Y.
\]
\QED

\paragraph{Remark}
The condition that $\bar p$ is injective is equivalent to
${\rm Im}\,p \bigcap {\rm coker}\,q=\{0\}$ (supposing that $p$ is injective).
If $pq=0$, then this condition is immediate since 
in this case ${\rm Im}\,p={\rm ker}\,q$.

\paragraph{} A dual way of solving the problem is
to define
\beq
\tilde q = qK : Y\to Z
\eeq
with
\beq
K= 1-pG_xp^* : Y\to Y.
\eeq
In this case, we get $\tilde q p =0$. We can then define
$\tilde\Delta_Z = \tilde q\tilde q^*$ and have a theorem analogous
to Thm.\ \ref{thm-bar}.

We summarize the previous results, and some more, in the following
\begin{Thm}
If $p$ is injective and $q$ is surjective, then
\begin{description}
\item{1.} 
$\bar\Delta_X$ is invertible if and only if $\tilde\Delta_Z$ is, or,
equivalently, if and only if ${\rm Im}\,p \bigcap {\rm coker}\,q
={\rm Im}\,q^*\bigcap{\rm coker}\,p^*=\{0\}$.
\item{2.} If any one is invertible then
\begin{description}
\item{a.} 
\[
\frac{\det\bar\Delta_X\,\det\Delta_Z}{\det\Delta_Y} =
\frac{\det\Delta_X\,\det\tilde\Delta_Z}{\det\Delta_Y} 
= 1;
\]
\item{b.} the operators
\[
\begin{array}{cccccccc}
\xi &=& p^*G_Yp &:& X&\to& X,\\
\zeta &=& qG_Yq^* &:& Z&\to& Z
\end{array}
\]
are identity operators;
\item{c.} the operators
\[
\begin{array}{cccccccc}
\hat\xi &=& p^*G_Yq^*qG_Yp &:& X&\to& X,\\
\hat\zeta &=& qG_Ypp^*G_Yq^* &:& Z&\to& Z
\end{array}
\]
are null operators.
\end{description}
\end{description}
\lab{thm-bartilde}
\end{Thm}
Notice that if $pq=0$ statements $2b.$ and $2c.$ 
follow trivially from
the commutativity of $p$ and $q$ with the Laplace operators.
The remarkable fact is that $\xi=\zeta=1$ and $\hat\xi= \hat\zeta=0$
even without this condition.

\paragraph{\em Proof}
To prove $1.$ we notice that
$\tilde\Delta_Z$ is invertible if and only if $\tilde q^*$ is injective,
i.e., if and only ${\rm Im}\,q^*\bigcap{\rm coker}\,p^*=\{0\}$
(supposing $q^*$ injective).
Since ${\rm Im}\,p={\rm coker}\,p^*$ and ${\rm Im}\,q^*={\rm coker}\,q$,
the last condition turns out to be 
${\rm Im}\,p \bigcap {\rm coker}\,q=\{0\}$
which is equivalent to the invertibility condition for $\bar\Delta_X$
(s.\ the preceding remark).

\subparagraph{}
Statement $2a.$ follows from Thm.\ \ref{thm-bar} after exchanging
$X$ with $Z$ and $p$ with $q^*$. By the way, either numerator is
equal to the determinant of the operator
\[
\left(\begin{array}{cc}
\Delta_Z & qp\\
p^*q^* & \Delta_X
\end{array}\right) : Z\otimes X\to Z\otimes X.
\]

To prove $2b.$, first of all we notice that we have the commutation rules:
\[
\begin{array}{lcl}
p\,\Delta_X - \Delta_Y\,q &=& - q^*qp,\\
q\,\Delta_Y - \Delta_Z\,q &=& qpp^*,
\end{array}
\]
together with their conjugates
\[
\begin{array}{lcl}
\Delta_X\,p^* - p^*\,\Delta_Y &=& - p^*q^*q,\\
\Delta_Y\,q^* - q^*\,\Delta_Z &=& pp^*q^*.
\end{array}
\]
Now we multiply the above relations by $G_*$ both from the left
and from the right getting
\beq
\begin{array}{lcl}
p\,G_X - G_Y\,q &=& G_Yq^*qpG_X,\\
q\,G_Y - G_Z\,q &=& -G_Zqpp^*G_Y,\\
G_X\,p^* - p^*\,G_Y &=& G_Xp^*q^*qG_Y,\\
G_Y\,q^* - q^*\,G_Z &=& -G_Ypp^*q^*G_Z.
\end{array}
\lab{pqcomm}
\eeq
Using the third relation of \rf{pqcomm}, we can write
\[
\xi = 1 - G_Xp^*q^*qG_Yp.
\]
Then we use the second relation of \rf{pqcomm} and obtain
\[
\xi = 1 - G_X R + G_X R\xi,
\]
where
\[
R = p^*q^*G_Zqp.
\]
Finally, we notice that
\[
\bar\Delta_X = \Delta_X - R.
\]
Applying $\Delta_X$ to the last formula for $\xi$, we get
\[
\bar\Delta_X \xi = \bar\Delta_X.
\]
If $\bar\Delta_X$ is invertible, this implies $\xi=1$.

A dual proof shows that $\zeta=1$ when $\tilde\Delta_Z$ is invertible.

To prove $2c.$, we use the fourth relation of \rf{pqcomm} and
obtain
\[
\hat\xi = (1-\xi)p^*q^*G_ZqG_Yp =0.
\]
A dual proof shows that $\hat\zeta=0$.
\QED

\paragraph{The infinite-dimensional case}
If $X$, $Y$ and $Z$ are (infinite-dimensional) Hilbert spaces, all the
previous theorems hold if we add the hypothesis that $p$ and $q$ have 
elliptic
Laplacians. In this case, by Hodge's theorem, the spectra of the Laplace
operators are discrete, the eigenspaces are finite-dimensional and
each Hilbert space is the direct sum of these eigenspaces. Therefore,
we can use the $\zeta$-function regularization. Namely, if we denote
by $\lambda$ the eigenvalues of a Laplace operator $\Delta$ and
by $d_\lambda$ the dimension of the corresponding eigenspace,
the $\zeta$-function is defined as
\[
\zeta_s(\Delta) = \Tr\Delta^{-s} = \sum_\lambda \lambda^{-s} d_\lambda
\]
for $s$ large enough and then analytically extended. The regularized
determinant is then defined as
\[
\det\Delta := \exp[-\zeta'_0(\Delta)].
\]

The proof of Thm.\ \ref{thm-qp0} can be refined by showing that
$p$ is an isomorphism between each eigenspace of $\Delta_X$ and
each eigenspace of $\Delta_{Y_1}$. 
This is essentially due to the fact that $p\Delta_X=\Delta_{Y_1}p$
and to the assumption that there are no zero eigenvalues. Therefore,
$\zeta_s(\Delta_X)=\zeta_s(\Delta_{Y_1})$ for all $s$.
By similar considerations
on $\Delta_{Y_2}$ and $\Delta_Z$, we get finally
\[
\zeta_s(\Delta_X) -\zeta_s(\Delta_Y)+\zeta_s(\Delta_Z)=0,
\quad\forall s.
\]
Deriving with respect to $s$ and setting $s=0$ yields (the logarithm of)
the required formula.

\subsection{Our case}
\lab{sec-ourc}
To apply the previous analysis to 
our case, we set
\[
\begin{array}{lcl}
X &=& \Omega^0(M,\ad P),\\
Y &=& \Omega^1(M,\ad P)\ominus \tHarm^1_A(M,\ad P),\\
Z &=& \Omplus.
\end{array} 
\]
The operators $p$ and $q$ correspond
to the operator $D_A$.
With these definitions we have
\[
\begin{array}{lcl}
\Delta_X &=& \Delta_A^{(0)},\\ 
\Delta_Y &=& \tDelta_A^{(1)},\\
\Delta_Z &=& \Delta_A^{(2,+)},
\end{array}
\]
Moreover, 
\[
\begin{array}{lcl}
\bar\Delta_X &=& \widehat\Delta_A^{(0)},\\
\xi &=& \tX_A,\\
\zeta &=& \tZ_A,\\
\hat\zeta &=& \widehat Z_A,
\end{array}
\]
where the operators on the r.h.s.\ are defined in \rf{defhDelta},
\rf{deftXZ} and \rf{defhZ}.

Recall that, by the definition of $\calN'$, 
$({\rm Im}\,D_A^*\bigcap{\rm coker}\,D_A^*)\bigcap\Omega^1(M,\ad P)=\{0\}$
if $A\in\calN'$. Therefore, by statement $1.$ of Thm.\ \ref{thm-bartilde},
$\widehat\Delta_A^{(0)}$ is invertible. 
Then, by statements $2b.$ and $2c.$ of Thm.\ \ref{thm-bartilde} we see that 
\beq
\tX_A = 1,\quad
\tZ_A = 1,\quad \widehat Z_A=0;
\lab{tXZ0}
\eeq
finally, by statement $2a.$, we have
\beq
J[A] = 1,
\lab{J1}
\eeq
with $J[A]$ defined in \rf{defJ}.

\end{document}